\documentstyle[11pt,a4wide,epsf]{article}
\def\ceff{{c_{\rm eff}}}
\def\cH{{\cal H}}
\def\cM{{\cal M}}
\def\cU{{\cal U}}

\def\ef#1,#2.{{\left(\frac{#1}{#2}\right)}}
\def\fe#1,#2.{{\left(-\frac{#1}{#2}\right)}}

\def\Mm #1,#2.{{\cM_{#1,#2}}}

\def\Ph#1,#2.{{\Phi_{(#1,#2)}}}

\def\Rm#1{\hbox{\rm\expandafter\uppercase\expandafter{\romannumeral #1}}}

\def\sm{\hbox{$S$--matrix}}
\def\sms{\hbox{$S$--matrices}}

\begingroup\makeatletter\ifx\SetFigFont\undefined
\def\x#1#2#3#4#5#6#7\relax{\def\x{#1#2#3#4#5#6}}%
\expandafter\x\fmtname xxxxxx\relax \def\y{splain}%
\ifx\x\y   % LaTeX or SliTeX?
\gdef\SetFigFont#1#2#3{%
  \ifnum #1<17\tiny\else \ifnum #1<20\small\else
  \ifnum #1<24\normalsize\else \ifnum #1<29\large\else
  \ifnum #1<34\Large\else \ifnum #1<41\LARGE\else
     \huge\fi\fi\fi\fi\fi\fi
  \csname #3\endcsname}%
\else
\gdef\SetFigFont#1#2#3{\begingroup
  \count@#1\relax \ifnum 25<\count@\count@25\fi
  \def\x{\endgroup\@setsize\SetFigFont{#2pt}}%
  \expandafter\x
    \csname \romannumeral\the\count@ pt\expandafter\endcsname
    \csname @\romannumeral\the\count@ pt\endcsname
  \csname #3\endcsname}%
\fi
\fi\endgroup

\makeatletter

\@addtoreset{equation}{section}
\def\section{\@startsection {section}{1}{\z@}{-3.5ex plus -1ex minus
 -.2ex}{2.3ex plus .2ex}{\large\bf\centering}}
\def\subsection{\@startsection{subsection}{2}{\z@}{-3.25ex plus%
 -1ex minus -.2ex}{1.5ex plus .2ex}{\bf}}
\def\subsubsection{\@startsection{subsubsection}{3}{\z@}{-3.25ex plus%
 -1ex minus -.2ex}{1.5ex plus .2ex}{\sl}}
\makeatother

\begin{document}
\parindent 12pt
\parskip 9pt

{
\parskip 0pt
\newpage
\begin{titlepage}
\begin{flushright}
%DAMTP--96--42\\
KCL--TH--96--7\\
hep-th/9605104\\
Revised November 26 1996\\[3cm]
\end{flushright}
\begin{center}
{\Large{\bf
On the relation between $\Phi_{(1,2)}$ and $\Phi_{(1,5)}$ perturbed\\
minimal models and unitarity.
}}\\[1.5cm]
{\large Horst Kausch%
\footnote{e-mail: hgk@mth.kcl.ac.uk}%
, G\'abor Tak\'acs%
\footnote{e-mail: takacs@hal9000.elte.hu}%
$\;$ and $\;$ G\'erard Watts%
\footnote{e-mail: gmtw@mth.kcl.ac.uk}%
}\\[8mm]
%{\em Department of Applied Mathematics and Theoretical Physics,}\\
%{\em Cambridge University,}\\
%{\em Silver Street, Cambridge, CB3 9EW, U.K.}\\[5mm]
${}^{1,3}$
{\em Department of Mathematics, King's College London,}\\
{\em Strand, London, WC2R 2LS, U.K.}\\[8mm]
${}^2$
{\em Institute for Theoretical Physics, E\"otv\"os University,}\\
{\em Puskin u. 5-7, H-1088 Budapest, Hungary}
\\[8mm] 
{\bf{ABSTRACT}}
\end{center}
\begin{quote}
We consider the RSOS \sms\ of the $\Ph1,5.$ perturbed minimal models
which have recently been found in the companion paper \cite{gt}. 
These \sms\ have some interesting properties, in particular, unitarity
may be broken in a stronger sense than seen before, while
one of the three classes of $\Ph1,5.$ perturbations (to be described)
shares the same Thermodynamic Bethe Ansatz as a related $\Ph1,2.$
perturbation. 
We test these new \sms\ by the standard Truncated Conformal Space
method, and further observe that in some cases the BA equations for
two particle energy levels may be continued to complex rapidity to
describe (a) single particle excitations and (b) complex eigenvalues
of the Hamiltonian corresponding to non-unitary \sm\ elements.
We make some comments on identities between characters in the two
related models following from the fact that the two perturbed theories
share the same breather sector.
\end{quote}
\vfill
\end{titlepage}
}

\section{Introduction}
\label{sec:one}
\setcounter{footnote}{0}

This paper is the companion to \cite{gt}, in which \sms\ were
conjectured for $\Ph1,5.$ perturbations of Virasoro minimal models by
RSOS reduction of the \sm\ of $a_2^{(2)}$ affine Toda theory. 
Throughout this paper we shall refer to equations in \cite{gt} as 
(1:nn).

The description of perturbed conformal field theories (PCFTs) as
affine Toda theories \cite{HMan} has provided much insight into their
structure --- the construction of the \sms\ for the particles in PCFTs
relies upon the existence of an affine quantum symmetry of the Toda
action \cite{BLec2} --- but there are still many interesting aspects
to investigate.
In this paper we consider the $a_2^{(2)}$ affine Toda theory at
imaginary coupling which can provide a model for both $\Ph1,2.$ and
$\Ph1,5.$ perturbations of minimal models. In the simplest case, the
$\Ph1,2.$ perturbation of the unitary Ising model (which has a minimal
$e_8$ \sm) is described by the same Toda Lagrangian as the $\Ph1,5.$
perturbation of the non-unitary $\Mm3,16.$ model. 

It is possible for the same Lagrangian to describe two such different
models because the \sms\ of a PCFT are obtained by `RSOS restriction'
of the Toda \sms\ which relies upon a $\cU_q(sl(2))$ subalgebra of the
quantum symmetry algebra, and that one may be able to choose more than
one such subalgebra leading to different \sms\ for the PCFTs. In this
way it turns out that the same Lagrangian theory may indeed describe
both a unitary and a non-unitary theory, and that even the particle
spectra and ground-state thermodynamics of these two theories may be
the same, yet the theories possess different \sms\, both being
consistent solutions of the \sm\ bootstrap equations for the particle
masses.

In the specific case of the $a_2^{(2)}$ Toda theory the RSOS
restriction leading to the $\Ph1,2.$ PCFTs has been performed in
\cite{smirnov,koubek,martins2} and that leading to the $\Ph1,5.$
perturbation in \cite{gt}. The way that the two perturbations which
arise from them are related falls into one of three classes; the
example considered by Martins in \cite{martins1} is in the first class
and we give the first examples of the other two classes in this paper,
together with a unified description.

The $a_2^{(2)}$ affine Toda theory appears to describe a non-unitary
theory, and while it is possible for the RSOS reduction of a
non-unitary theory to yield a unitary scattering theory (as with the
$\Ph1,2.$ perturbations of unitary minimal models) for most reductions
the scattering theory is also non-unitary. However, for all the
reductions proposed so far, this has been a rather weak form of
non-unitarity, with the Hamiltonian having real eigenvalues bounded
below, and simply the Hilbert space having an indefinite metric.
However, in some of our models, we show how unitarity is violated in a
stronger sense, and the Hamiltonian of the reduced theory may have
complex eigenvalues as well.

The outline of the paper is as follows: in 
\ref{sec:cft} we describe briefly the minimal models of conformal
field theory and their perturbations and the way in which two different
perturbed conformal minimal models, one perturbed by the $\Ph 1,2.$
field and one by the $\Ph 1,5.$ field, may be related to the same
affine Toda Lagrangian field theory, and how the BRS reduction of the
two different underlying Liouville theories naturally leads to two
different \sms\ for the two perturbed minimal models.

We then pass to tests of the predicted \sms\ in the case of the
$\Ph 1,5.$ perturbations. In section \ref{sec:tba} we recall the 
predictions of the Thermodynamic Bethe Ansatz for the ground state
energy of the particle system on a circle, which leads to predictions
for the effective central charge, the mass-gaps and of the coefficient
in the linear term in the asymptotic behaviour of the ground-state
itself. We also show how in many cases the two perturbed minimal
models which can be derived from Smirnov's \sm\ give the same TBA
system, while in other cases this is not so.  We also recall how the
higher multi-particle energy levels can be predicted using the
ordinary Bethe ansatz.

The first prediction can be checked analytically while the other three
can be checked using the Truncated Conformal Space Approach (TCSA) of
Yurov and Al.~Zamolodchikov \cite{tcsa1}. 
In section \ref{sec:tcsa} we use this method to show that our
predictions of the particle spectra and \sms\ are correct in three
models, the minimal models $\Mm3,10.$, $\Mm3,14.$ and $\Mm3,16.$
perturbed by the field $\Ph1,5.$. 

Furthermore we show that the Bethe Ansatz can be usefully continued to
complex rapidities in two different cases. Firstly, we show that 
the ordinary Bethe Ansatz may also be a good description of a single
particle state, when continued to imaginary rapidity. This presumably
indicates that the Bethe Ansatz equations provide the dominant
corrections to the full TBA equations for single particle states
developed in \cite{fullTBA}. Secondly, we show that for the
non-unitary \sms\ the Bethe Ansatz still provides a good description
of the complex eigenvalues analogous to two-particle states
found in the TCSA investigation.

In section \ref{sec:ad} we give some comments on the possible ways
in which the \sm\ may be realised and the connection with the
different partition functions and periodicity conditions which may
occur for the perturbed models, and in section \ref{sec:chid} we list
some relations between the Virasoro characters of related minimal
models. Finally we give our conclusions in section~\ref{sec:conc}. 

We give some details of the structure constants in the
conformal field theories we use in appendix \ref{app:cc}.

\section{Minimal models and their perturbations}
\label{sec:cft}

A conformal field theory is characterised by the central charge $c$ of
the Virasoro algebra, the sets of conformal weights $\{h,\bar h\}$ of
the primary fields
$\Phi_a(z,\bar z) \sim \phi_h(z)\phi_{\bar h}(\bar z)$
and the three point couplings of these fields 
$\langle \Phi_a | \Phi_b(1) | \Phi_c \rangle$.
The minimal models of conformal field theory were first
described in \cite{bpz}, and we denote them by $\Mm r,s.$ where $r,s$
are coprime integers greater than 1. 
The Virasoro central charge in these models take the values 
\begin{equation}
  c = 1 - \frac{6(r-s)^2}{rs}
\ ,
\end{equation}
and in $\Mm r,s.$ there are $(r-1)(s-1)/2$ possible values of $h$ and
$\bar h$, given by  
\begin{equation}
  h_{m,n} = \frac{ (ms - nr)^2 - (r-s)^2 }{4 rs }
\ ,\ \ 
  0<m<r,\ 0<n<s
\ ,
\end{equation}
although which actual pairs $\{h,\bar h\}$ are the conformal weights
of fields is not determined a priori.
For $|r-s|=1$ $\Mm r,s.$ is unitary, and the lowest value of $h$ is
$h_{1,1} = 0$; in all other cases there is some allowed value of
$h<0$, but there is a unified formula for the effective central charge
$c - 24 h_{min}$, 
\begin{equation}
  \ceff 
\equiv c - 24 h_{min}
= 1 - \frac{6}{rs}
\,.
\end{equation}
One can consider perturbations of these conformally invariant field
theories by the addition to the action of some term
\begin{equation}
  \lambda 
  \int {\rm d}^2x \ \ 
  \Ph m,n.(x)
\ ,
\end{equation}
where $\Ph m,n.$ is the conformal primary field corresponding to
$h=\bar h= h_{m,n}$. 
First-order perturbation theory calculations, backed up
by arguments based on counting of states, suggest that the
perturbed field theory is integrable for $(m,n)$ one of 
$(1,2)$, $(1,3)$, $(1,5)$, $(2,1)$, $(3,1)$ and $(5,1)$. For the
unitary minimal models $\Mm r,r+1.$, only $(1,2), (1,3)$ and $(2,1)$
are relevant perturbations, and consequently these have attracted the
most attention to date. 

Much insight can be gained into minimal models and perturbed minimal
models by considering particular realisations as Lagrangian field
theories.
The Liouville action for a scalar field $\phi$,
\begin{equation}
  S_0
= \int {\rm d}^2 z \ \ \left( \ 
  \frac{1}{4\pi}\partial_z \phi \partial_{\bar z} \phi+ 
  \exp(\beta \sqrt 2 \phi )
  \ \right)
\ ,
\label{eq:sl}
\end{equation}
provides a realisation of the minimal models with 
$c = 13 + 6 \beta^2 + 6 \beta^{-2}$. The minimal models correspond to
the purely imaginary values of $\beta = i \sqrt{r/s} $. Although this
means that care must be taken with the interpretation of the action,
Felder has provided a construction \cite{Feld1} which shows that the
correlation functions defined in this theory are indeed those of the
minimal models as found by Dotsenko and Fateev \cite{DFat}: for the
vacuum sector one must not consider the full Hilbert space of a scalar
field but instead the reduced space 
${\rm Ker}\ Q /{\rm Im}\ :Q^{r-1}:$ where $Q$ is the screening charge
\begin{equation}
Q = \int {\rm d}z\ \exp( \sqrt 2 \phi / \beta) 
\ ,
\label{eq:sc}
\end{equation}
and $:Q^{r-1}:$ is an appropriately defined normal-ordered multiple
integral. 
The vertex operators  
\begin{equation}
  \exp( - \alpha_{m,n} \phi /\sqrt 2 )
\ ,\ \ 
  \alpha_{m,n} = (n-1) \beta - (m-1)/\beta
\ ,
\end{equation}
transform as primary fields with conformal weight $h_{m,n}$.
Consequently, one can construct a formal perturbed conformal field
theory action as 
\begin{equation}
  S = 
  S_0
+ \lambda \int {\rm d}^2 z\ \exp( -  \alpha_{m,n} \phi/\sqrt 2)
\ .
\end{equation}
For $(m,n)$ = $(1,3)$ this gives the standard sine-Gordon action,
whereas for $(m,n)=(1,2)$ or $(1,5)$ we obtain the Bullough-Dodd or
Zhiber-Mikhailov-Shabat (ZMS) model, which is the affine Toda theory
related to $a_2^{(2)}$.  

It is this latter model in which we are most interested, and to fix
on a normalisation of the coupling constant, we shall say that the ZMS
model with coupling constant $\gamma$ has the action
%%%%%%%%%%%%%%%%%%%%%%%%
%
% NB gamma rescaled by pi
%
\def\newg{{\gamma}}
\def\oldg{{\gamma/\pi}}
\def\ourg{\newg}
%
%%%%%%%%%%%%%%%%%%%%%%%
\begin{equation} 
  S_{ZMS}(\ourg)
= \int {\rm d}^2 x
  \left((\partial_\mu\varphi)^2
             + \frac{m^2}{\ourg}
       \left(  \exp( i\sqrt{8\ourg}\varphi)
             + 2 \exp(-i\sqrt{2\ourg}\varphi)\right)
       \right)
\ ,
\label{zmslagr}
\end{equation}
where $m$ is a coupling constant with dimension of mass. The connection 
between the variables in (\ref{eq:sl}) and (\ref{zmslagr}) is given by
\begin{equation}
\phi=\sqrt{4\pi}\varphi\ , \ \beta=i\sqrt{\frac{\gamma}{\pi}}\ .
\end{equation}

It is immediately apparent that the ZMS action may be viewed as a
perturbed conformal field theory in two different ways:  each of the
two exponentials may be thought of as the perturbation and the
remainder as the Liouville action: 
\begin{equation}
  S_{ZMS}(\ourg) 
= \cases{
  \Mm r,s.   + \Ph 1,2.   & $ \ourg = \pi r /s $ \cr
  \Mm r',s'. + \Ph 1,5.   & $~\ourg = 4\pi r'/s'$ \cr}
\ .
\label{eq:2acts}
\end{equation}
The actual values of $r',s'$ depend on the arithmetic properties of
$r$ and $s$, since we require $r',s'$ coprime.
Hence the three distinct cases of table \ref{tab:rs} can arise.

\begin{table}[htb]
\[
\begin{array}{c|c|cc}
      &  r \hbox{ mod } 4 & r'  &   s' \\
\hline
\Rm 1 &                0  & r/4 &   s  \\
\Rm 2 &                2  & r/2 &  2s  \\
\Rm 3 &               1,3 & r   &  4s  \\
\end{array}
\]
\caption{Relation between $(r,s)$ and $(r',s')$}
\label{tab:rs}
\end{table}

These three cases differ in the way the effective central charges of
$\Mm r,s.$ and $\Mm r',s'.$ are related.
There are interesting examples of each of these cases as follows:

\begin{itemize}

\item[\Rm 1] 

Here $\ceff = 1 - 6/rs$, $\ceff' = 1 - 24/(rs)$ so that
$\ceff' < \ceff$. Since the effective central charge is a measure
of the number of degrees of freedom in the UV theory, we expect that
there will be fewer particles in $\Mm r',s'.+ \Ph 1,5.$ than in
$\Mm r,s.+\Ph 1,2.$.

The only $(1,5)$ perturbation that has been investigated before falls
into this class, being $\Mm 2,9.+\Ph 1,5.$ which is related
to $\Mm 8,9.+\Ph 1,2.$ and which was discussed by Martins et al in
\cite{martins1,martins2}. 

\item[\Rm 2]

Here $\ceff' = \ceff$ and we expect the greatest similarity
between the two models.

We will discuss two models,
$\Mm 6,5. + \Ph 1,2.$ which is related to $\Mm 3,10.+\Ph 1,5.$, and
$\Mm 6,7. + \Ph 1,2.$ which is related to $\Mm 3,14.+\Ph 1,5.$. 

The first of these two models is unusual in that it can be expressed
as a tensor product, $\Mm3,10. \equiv \Mm2,5.\otimes\Mm2,5.$. The
fields in $\Mm 3,10.$ can be classified into ${\rm Z}_2$-even and $\rm
Z_2$-odd fields with respect to the ${\rm Z}_2$ map provided by flipping
the tensor product. The perturbing operator  
$\Ph 1,5.$ is in the ${\rm Z}_2$-even sector and is nothing other than
$(1\otimes\Phi_{(1,2)} +\Phi_{(1,2)} \otimes 1)/\sqrt{2}$.
As a result, 
\begin{equation}
  \Mm 3,10.+\Ph 1,5. 
= \left(\Mm 2,5. + \Ph1,2.\right)^{\otimes 2}
\ ,
\label{eq:310=25}
\end{equation}
is identically true in the ${\rm Z}_2$ even sector, and the spectrum
consists of two self-conjugate particles of equal mass, each with the
\sm\ of the $\Mm2,5.+\Ph1,2.$ model.

In the second case the spectrum consists of six particles 
$\{l,\bar l,L,h,\bar h,H\}$ for $\Mm6,7.+\Ph1,2.$ and
$\{A,\bar A,C,B,\bar B,D\}$ for $\Mm3,14.+\Ph1,5.$
The \sm\ in the second case is unusual in that it is not
possible to define $S_{AB}$ and $S_{A\bar B}$ satisfying the usual
unitarity condition, i.e. they are not pure phases. However, a
formal application of the \sm -bootstrap gives
$S_{AB} = S^{KL}_1, S_{A\bar B} = S^{KL}_2$ as in (1:79), and 
the remaining \sms\ can be found by application of the \sm -bootstrap
to the poles in $S_{AB}, S_{A\bar B}$.

\item[\Rm 3]

In this case $\ceff' > \ceff$ and we expect there to be more
particles in $\Mm r',s'.+\Ph 1,5.$ than in $\Mm r,s.+\Ph 1,2.$.

Here we consider the case of the thermal perturbation of the Ising
model, i.e. $\Mm 3,4.+\Ph 1,2.$ for which the corresponding
model is $\Mm 3,16.+\Ph 1,5.$.

Although the bootstrap was not completed in [1], again we expect that
$\Mm3,16.+\Ph1,5.$ has the same peculiarities as $\Mm3,14.+\Ph1,5.$.

\end{itemize}

% The \sms\ of the example of type \Rm 1 are already known -- the case
% $\Mm8,9. + \Ph1,2.$ is among those considered in \cite{smirnov}, and
% the \sms\ of the related model $\Mm2,9.+\Ph1,5.$ are found,
% andchecked against perturbation theory in \cite{martins1}. However,
% in \cite{martins2} the \sms\ of $\Mm2,9.+\Ph 1,5.$ were found by an
% adhoc method, whereas they can be found easily from the general
% scheme given in \cite{gt} in which the three examples of types \Rm 2
% and  \Rm 3 can also be found. Firstly we recall the expressions for
% the \sms\ as found there. 

\section{TBA predictions from the \sms.}
\label{sec:tba}

As is well-known, the method of Thermodynamic Bethe Ansatz (TBA) is a
powerful method for extracting the finite-size behaviour of a massive
theory on a circle. 
One can also calculate this behaviour for the perturbed conformal
field theory, both in perturbation theory about the conformal field
theory and also numerically, and together these two methods provide a
stringent test of the conjectured \sms.

We first formulate the conformal field theory on a cylinder of size
$R$, and consider the Hamiltonian which propagates along the cylinder:
\begin{equation} 
  H
= H_{CFT}
 +\lambda\int {\rm d}x\ \Phi(z,{\bar z})\ ,
\end{equation}
where 
\begin{equation}
  H_{CFT}
= \frac{2\pi}{R}\left(L_0+{\bar L}_0-\frac{c}{12}\right)
\ ,
\end{equation}
is the Hamiltonian of the ultraviolet CFT, $\Phi(z,{\bar z})$ 
is some relevant operator corresponding to the perturbation and 
$\lambda$ is a dimensionful coupling. If the conformal dimension of the 
perturbing field $\Phi(z,{\bar z})$ is $d_{\Phi}=2-y$, then $\lambda$ has 
mass dimension $y$. In terms of a specific mass unit $m$ (which will 
be taken to be the mass of the lowest lying excitation of the model) 
we have
\begin{equation} 
\lambda = \alpha m^y\ ,
\end{equation}
where $\alpha$ is a dimensionless constant. The perturbing Hamiltonian
is given by the second term:
\begin{equation} 
  V
= \lambda\int {\rm d}x\ \Phi(z,{\bar z})\ ,
\end{equation}
Perturbation theory in $\lambda$ gives the following expansion 
for the ground state energy
\begin{equation} 
  E_0(R)
= -\frac{\pi \ceff}{6R}
  -\frac{2\pi}{R} \sum\limits_{n=1}^{\infty}C_n(R^y\lambda )^n
\ .
\label{eq:e0pcft}
\end{equation}
The terms in the series 
come from the perturbative corrections, with coefficients
\begin{equation} 
  C_n
= \frac{(-1)^n}{n!}R^{2-ny}\int\langle \Phi (0)
  \prod\limits_{j=1}^{n-1}
  \Phi (\xi_j)\ {\rm d}^2\xi_j\ \rangle_{0,c}\ ,
\end{equation}
where the integration is taken over the cylinder and the indices $0,c$ 
refer to the connected contribution to the Green function in the 
original conformal field theory on the cylinder. 
For unitary theories, $C_1$ is zero, while in the case of
non-unitary models the following formula is obtained in \cite{tba3},
\begin{equation} 
  C_1
= -(2\pi )^{1-y}C_{\Phi_0\Phi\Phi_0}\ ,
\end{equation}
where 
$ 
  C_{\Phi_0\Phi\Phi_0} 
$
is the operator product coefficient for the perturbing operator $\Phi$ 
and the operator with the lowest (negative) conformal weight $\Phi_0$.

{}From the TBA, the same expansion can be computed using the
conjectured scattering amplitudes. If we introduce the scaling
length $r=mR$, where $m$ is taken to be the mass of the lightest
particle in the model, then the TBA yields $E_0(R)$ as the series
\begin{equation} 
   \frac{E_0(R)}{m}
= -\frac{\pi{\tilde c}}{6r}-Ar-\frac{2\pi}{r}
   \sum\limits_{n=1}^{\infty}a_n r^{yn}
\ .
\label{eq:e0tba}
\end{equation}
The first two terms, $\tilde c$ and $A$, can be calculated
analytically from the TBA whereas the $a_n$ have to be calculated
numerically.  
The agreement of the two series expansions (\ref{eq:e0pcft}) and
(\ref{eq:e0tba}) leads to the following relations:
\begin{equation} 
  \ceff
= {\tilde c},\  
  a_n
= C_n(m^y\lambda)^n\ .
\label{compare}\end{equation}
The linear term in (\ref{eq:e0tba}), which is absent in (\ref{eq:e0pcft}), 
is interpreted as the bulk energy term in finite volume.
 
The second relation in (\ref{compare}) yields a mass gap formula 
in the special case $n=1$,  
\begin{equation} 
  \lambda
= \frac{a_1}{C_1}m^y\ .
\end{equation}
The first nontrivial check is provided by the reproduction of the 
correct effective central charge, while the bulk energy constant and
the mass-gap can be measured using TCSA.
The $n>1$ equalities could be used to perform further consistency
checks, but we shall not do this.

Now we give the TBA results for the models
$\Mm 3,10.$ and $\Mm 3,14.$ by relating them to known results.

\subsection{TBA  for $\Mm3,10.+\Ph1,5.$}

It was pointed out in \cite{tba1} that the TBA calculation for the
scaling Yang-Lee model (which is $\Mm 2,5.+\Ph 1,2.$) is identical to
the TBA calculation for the scaling Potts model 
$\Mm 5,6.+\Ph 2,1.$ for the following reason:
The scaling Yang-Lee model contains one scalar particle $B$ with the 
\sm\ 
\begin{equation} 
  S_{BB}
= \left(\frac{1}{3}\right)\left(\frac{2}{3}\right)\ ,
\end{equation}
while the spectrum of the scaling Potts model consists of a 
conjugate particle pair $A$ and $\bar A$ with the \sm\
\begin{equation} 
  S_{AA}
= \left(\frac{2}{3}\right)\ ,\  
  S_{A\bar A}
= \left(\frac{1}{3}\right)\ .
\end{equation}
The two \sms\ satisfy
\begin{equation} 
S_{AA}S_{A\bar A}=S_{BB}\ .
\end{equation}
As shown in \cite{tba1}, this means that the TBA equation in the charge 
symmetric Gibbs state for the scaling Potts model is equivalent to two 
copies of the scaling Yang-Lee model. 

However, the model $\Mm 3,10.+\Ph 1,5.$ is explicitly equal to two
copies of the Yang-Lee model, and therefore the ground  state energy
computed from the TBA of the model $\Mm 5,6.+\Ph 2,1.$ is identical to
that of the model $\Mm 3,10.+\Ph 1,5.$. 

\subsection{TBA  for $\Mm3,14.+\Ph1,5.$}

The model $\Mm 3,14.+\Ph 1,5.$ is related to $\Mm 6,7.+\Ph 1,2.$. 
The \sm\ of the latter is the so-called minimal $e_6$ \sm, derived
in \cite{E6}. 
%The mass spectra of the two models are identical, and so
%we shall use the same labels for the particles in the two theories.
If we denote the \sm\ of the $i$ and $j$ particles in the non-unitary
model by $S_{ij}^{(3,14)}\ $, while that of the corresponding
particles in the unitary case by
$S_{ij}^{(6,7)}$, then we have the following interesting relation,
\begin{equation} 
S_{ij}^{(3,14)}S_{i{\bar\jmath}}^{(3,14)}=
S_{ij}^{(6,7)}S_{i{\bar\jmath}}^{(6,7)}
\ ,
\label{eq:smateq}
\end{equation} 
even for the case when $S^{3,14}_{ij}$ is not a pure phase,
\begin{equation}
S_{AB}\,S_{A\bar B} = S_{lh}\,S_{l\bar h}
\ .
\end{equation}
This implies that the ground state energy calculated using the 
charge symmetric Gibbs state for TBA is the same in the two models.

The TBA for $\Mm6,7.+\Ph1,2.$ has been analysed in \cite{tba3}, where
they find
\begin{equation}
  \tilde c = 6/7\ ,\  \
  A^{TBA}  = \frac{1}{6+2\sqrt{3}}\ ,\ \      % 0.105662433
  a_1      = 0.0027765\ldots\ ,               % 0.00277651827
\end{equation}
For $\Mm3,14.$, $c=-114/7$ and the operator $\Phi_0$ with the minimal
dimension is $\Phi_{(1,5)}$ with $h_{min}=-5/7$, so that
$\ceff=6/7$ as predicted.
Using $C_{\Phi\Phi_0\Phi}=-20.3634\ldots $,   % -20.3634054
the mass-gap $m$ is given by 
\begin{equation}
  \lambda = \alpha\, m^{24/7}\ ,\ \
  \alpha = -0.011833\ldots\ .                  % -.0118327844
\end{equation}

\subsection{TBA for $\Mm3,16.+\Ph1,5.$}

The arguments used in the two previous subsections to show that the
TBA analysis of the two related models are equal are not valid here
since the non-unitary model contains more particles in its spectrum
than the corresponding unitary one. Therefore one does not expect the
TBA prediction derived from the Ising model results to hold in this
case.  

For example, consider the effective central charge:
In the $\Mm 3,16.+\Ph 1,5.$ model we have $c=-161/8$, $h_{min}=-7/8$
(corresponding to the operator $\Ph 1,5.$) and 
so $\ceff=7/8$.
The related model is $\Mm3,4.+\Ph1,2.$ with $\ceff = 1/2$, and which
has 8 particles with the minimal $e_8$ \sm, for which $\tilde c = 1/2$
as required. 

In the $e_8$ TBA \cite{tba3}, the mass gap is 
expressed in terms of the lowest lying breather. If we try to use the
$e_8$ TBA expansion for $\Mm 3,16.$ given in \cite{tba3}, 
we get 
\begin{equation} 
               a_1 = -0.001236\ldots\ ,\ \   % -0.00123610723
 C_{\Phi\Phi_0\Phi}=  51.85\ldots\ ,         %      51.8472202
\end{equation}
and therefore the predicted mass gap is
\begin{equation} 
  \lambda = \alpha m_{B_1}^{15/4}\ ,\ \ 
  \alpha  = -.003735\ldots                  % -.00373529847
\end{equation}
The bulk energy constant can be obtained from \cite{tba3} as follows:
\begin{equation} 
  A^{TBA}
= \frac{1}
       {4\left(\sin\frac{\pi}{15}+
               \sin\frac{\pi}{3}+
               \sin\frac{2\pi}{5}+
               \sin\frac{3\pi}{5}+
               \sin\frac{2\pi}{3}+
               \sin\frac{14\pi}{15}
          \right)}
= .06173\ldots                              % .0617285898\ .
\end{equation}
We give these simply to show that they do not agree with the numerical
results of the TCSA in section \ref{sec:tcsa}, as indeed they should
not. 

\subsection{BA predictions from the \sm}

The TBA system is a set of complicated non-linear integral equations
for the ground-state energy $E_0(R)$, and even more complicated
equations for higher energy levels \cite{fullTBA}. There is a much
simpler set of equations which give the dominant behaviour of
two-particle excitations, which comes simply from the usual Bethe
Ansatz for a two-particle wave function as follows.

On a circle the momenta of two-particle states are quantised. If we
consider the zero momentum sector with particles $i$ and $j$ of masses
$m_i$, $m_j$, momenta $p$, $-p$, and total energy $E(R)$, then we have
\begin{equation}
  p = m_i \sinh( \theta_1 ) = - m_j \sinh( \theta_2 )
\ ,\ \ 
  E(R)-E_0(R) = \sqrt{ p^2 + m_i^2 } + \sqrt{ p^2 + m_j^2}
\ ,
\label{eq:ba1}
\end{equation}
where $E_0(R)$ is the ground state energy. If the particles have
purely elastic scattering with \sm\ $S_{ij}(\theta)$ then
the quantisation condition is  
\begin{equation}
  \exp\left( i p R \right)\ S_{ij}(\theta_1 - \theta_2)\ =\ \pm 1
\ ,
\label{eq:ba2}
\end{equation}
with a $-1$ for $i$ and $j$ both fermions and $1$ otherwise.
Consequently it is possible to extract the \sm, or more properly the
phase shift $\delta(\theta) = -i \log(S(\theta))$,  from the function
$E(R)-E_0(R)$. If there are two or more particles of the same mass
with non-diagonal scattering then the quantisation condition is
(\ref{eq:ba2}) with $S_{ij}(\theta)$ replaced by the eigenvalues of the
two-particle \sm.
We are able to identify several such two-particle lines in our
numerical results and we can compare our conjectured \sms\ with those
of the related models.  

However, there is an ambiguity in our measurements of $S(\theta)$
which we should mention. The particle statistics are determined by the
sign $\pm1$ in equation (\ref{eq:ba2}); there is also the notion of
`type' which enters the TBA and which is given by $(\pm1) S(0)$. The
folklore is that all particles in theories such as ours are of
fermionic type, that is $(\pm 1)S(0) = -1$, and this is the convention
we have used in our TBA and BA calculations. However, the TBA and BA
predictions are unaltered by $S(\theta)\to-S(\theta)$ and
fermion$\leftrightarrow$boson, and this ambiguity can only be fixed by
requiring the correct signs in the \sm\ bootstrap.

\subsection{Extension of BA predictions from the \sm.}
\label{ssec:BA2}

There are two situations in which we would like to consider extending
equations (\ref{eq:ba1}) and (\ref{eq:ba2}) to complex rapidities. 

Firstly, if the \sm\ is unitary, i.e. a pure phase for real rapidity,
then we can reasonably expect $E(R)$ to be real. This
requires that either $\theta_i$ are real and $p$ is real, or
$\theta_i$ is purely imaginary and and $p$ is purely imaginary. In the
second case, the energy will be less than the sum of the particle
masses, and so cannot correspond to an excited state of these two
particles. However, if one considers the full TBA equations for the
single particle state $k$, then it is quite possible that for large
$R$ that they are dominated by this BA equation, if $k$ appears as a
bound state pole in $S_{ij}$. Although this may give the leading
corrections for large $R$, as is the case for the Yang-Lee model%
\footnote{One can check that the leading correction to the
single-particle mass found in \cite{tcsa1} is exactly given by the BA
equation at imaginary rapidity}
to the $\Mm2,5.+\Ph1,2.$, typically this will not be a good
approximation for small $R$ as other corrections become dominant.
However, we shall see that this need not be the case.

Secondly, if the \sm\ is not a pure phase for real rapidity
difference, then requiring $E(R)$ real will lead to no solution of
(\ref{eq:ba2}). However, in this case, we can instead demand that $R$
is real, which is certainly a requirement, and which will typically
lead  to complex values of $\theta_i$ and $E(R)$. We have the first
such occurrence in $\Mm3,14.+\Ph1,5.$, where 
$S_{AB}$ and $S_{A\bar B}$ are not pure phases. They are
exchanged under crossing, and as a result, the values of $E(R)$ which
we find by applying the BA to these two \sms\ are conjugate.
The BA is derived under the assumption that the particles are well
separated and with real momenta. which is certainly not the case here;
however it is remarkable that this simple rule does indeed describe
the spectrum very well, and we compare these values with the TCSA data
in subsection \ref{ssec:314}.

\section{Numerical analysis in the Truncated Conformal Space Approach}
\label{sec:tcsa}

The Truncated Conformal Space Approach (TCSA) consists of calculating
the matrix elements of (\ref{eq:tcsah}) exactly for a finite
dimensional subspace $\cH$ of the full Hilbert space, and then
numerically diagonalising the resulting matrix, as explained in
\cite{tcsa2}.   Following the discussion in \cite{tcsa2}, the spectrum
of the truncated Hamiltonian $h_{TCSA}$ exhibits three distinct
scaling regimes; $r$ small in which the UV conformal field theory
dominates, $r$ large in which the truncation effects dominate and an
intermediate region in which one hopes that the spectrum will
approximate that of the massive theory on a circle.
The TCSA method is fully described in e.g.\ \cite{tcsa2}, but we
recall the main elements here. 
The full Hamiltonian is:
\begin{equation}
  h(r)
= \frac{2\pi}r\left( H_0 + \alpha\frac{r^y}{(2\pi )^{y-1}}H_1\right)
\ ,
\label{eq:tcsah}
\end{equation}
in units in which $m=1$ and where 
\begin{equation}
  H_0
= L_0+{\bar L}_0-\frac{c}{12}\ .
\end{equation}
The matrix elements of the perturbing Hamiltonian can be evaluated 
using the formula
\begin{equation} 
  \langle \Psi_f | V |\Psi_i\rangle 
= \frac{2\pi}{R}\lambda \frac{R^y}{(2\pi )^{y-1}} 
  \langle \Psi_f | \Phi (1)|\Psi_i\rangle_{\rm plane} \ ,
\end{equation}
where the matrix element of the perturbing field is taken on the 
conformal plane and evaluated using standard conformal field theory
techniques. From this the operator $H_1$ can be identified with the
perturbing field $\Phi (1)$ itself.  

There are some symmetries which make the calculation of $h_{TCSA}$
simpler: firstly, since the Hamiltonian is Lorentz invariant, we can
restrict attention to the spin-zero sector of the full Hilbert space.
Secondly, in our models the perturbing operator $\Phi(1) = \Ph1,5.(1)$
preserves  distinct sectors of the full Hilbert space, and as a result
we may diagonalise $h_{TCSA}$ on these sectors separately. We can also
choose to normalise the highest weight states as  
\begin{equation} 
  \langle \Phi_j |\Phi_i\rangle
= \pm\delta_{ij},
\end{equation}
where the signs are chosen to ensure the reality of the three-point 
couplings, which are themselves given in appendix \ref{app:cc}.

As discussed in section \ref{sec:tba}, the following numerical results
can be extracted from the TCSA: 
(1) The ground-state energy: in the scaling regime the ground state
energy should be approximately $ E_0(r)/m \simeq A r$, and one can
measure $A$. (2) The mass-gap can be found and compared with the TBA
predictions. In practice, we adopt the value of $\alpha$ from
the TBA and then we expect that the lightest particle has mass 1.
(3) Some of the two-particle lines can be identified, and for those,
the two-particle energies can be compared with the Bethe Ansatz
results derived from the \sm.
We now present the results of the Truncated Conformal Space
calculations for the two models $\Mm3,14.+\Ph1,5.$ and
$\Mm3,16.+\Ph1,5.$.

\subsection{The TCSA results for $\Mm3,14.+\Ph1,5.$}
\label{ssec:314}

The allowed values of $h$ are $h_{1,i},\ 0<i<14$. It is possible to
consider several different sectors in which $\Ph1,5.$ acts locally but
which are not themselves necessarily mutually local. In particular, we
can consider the three sectors `even', `odd' and `twisted' for which
the fields all have integer spin, with the following sets of 
$\{h_{1,i},\bar h_{1,j}\} \equiv [i,j]$, 
\begin{equation}
\begin{array}{ll}
\hbox{Sector}\ \ &\hbox{Field content} \\
\hbox{even}       & [1,1],\, [3,3],\, [5,5],\, [7,7],\, 
                   [9,9],\, [11,11],\, [13,13] \\
\hbox{odd}      & [2,2],\, [4,4],\, [6,6],\, [8,8],\, 
                   [10,10],\, [12,12] \\
\hbox{twisted}   & [1,13],\, [3,11],\, [5,9],\, [7,7],\, 
                   [9,5],\, [11,3],\, [13,1] \\
\end{array}
\end{equation}
The coupling constants for first two sectors are those of the `A'
type modular invariant, which we give in appendix \ref{app:cc},
the even\ and twisted sectors correspond to the `D' type modular
invariant. 

To establish that this model has the spectrum we have predicted, it
is only necessary to consider the `even' sector.
We truncated the space by requiring $L_0 + \bar L_0 \leq 10$
which leaves 285 states. In figure \ref{fig:314.1} we give the real
eigenvalues of $h(r)$ and pick out the first 13 lines; it is clear
that scaling has set in for the first three eigenvalues by $r=10$.  

{
\begin{figure}[ht]
\[
\begin{array}{ll}
\epsfxsize=7.4cm
\epsfbox[0 0 288 288]{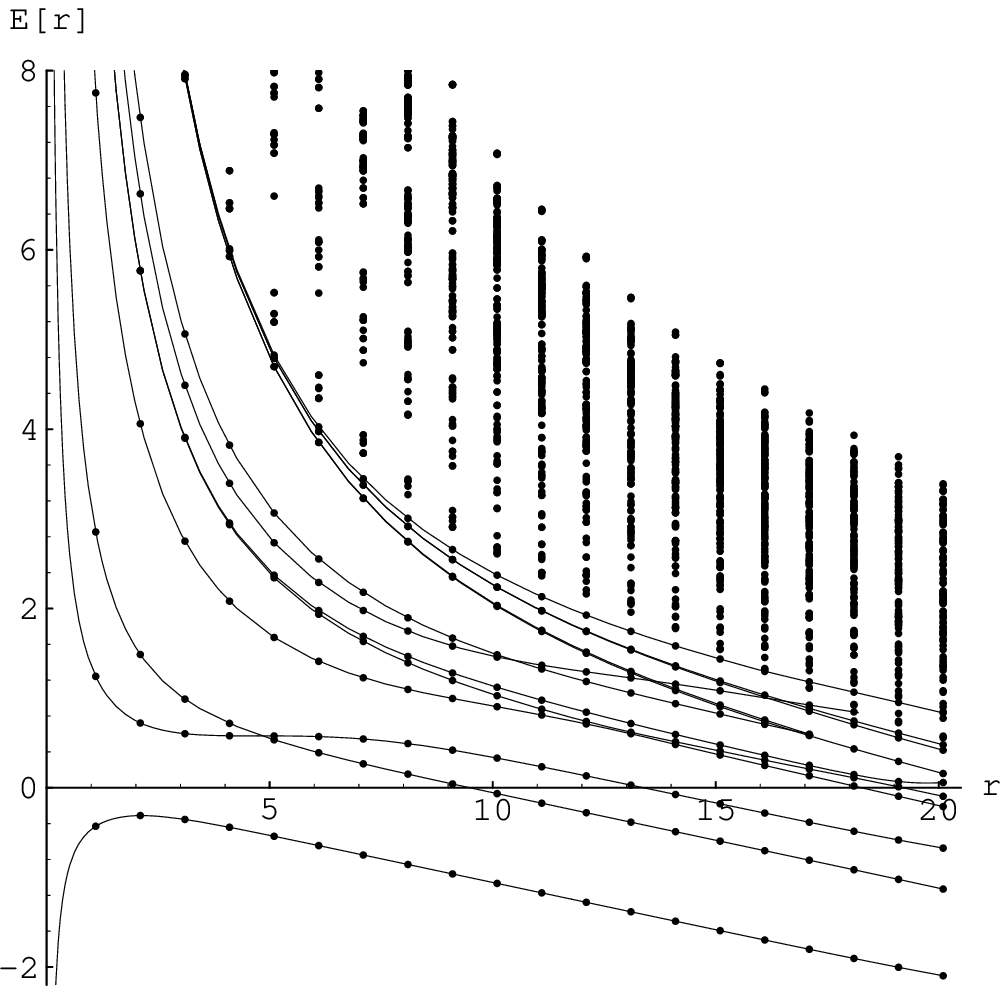} &
\epsfxsize=7.4cm
\epsfbox[0 0 288 288]{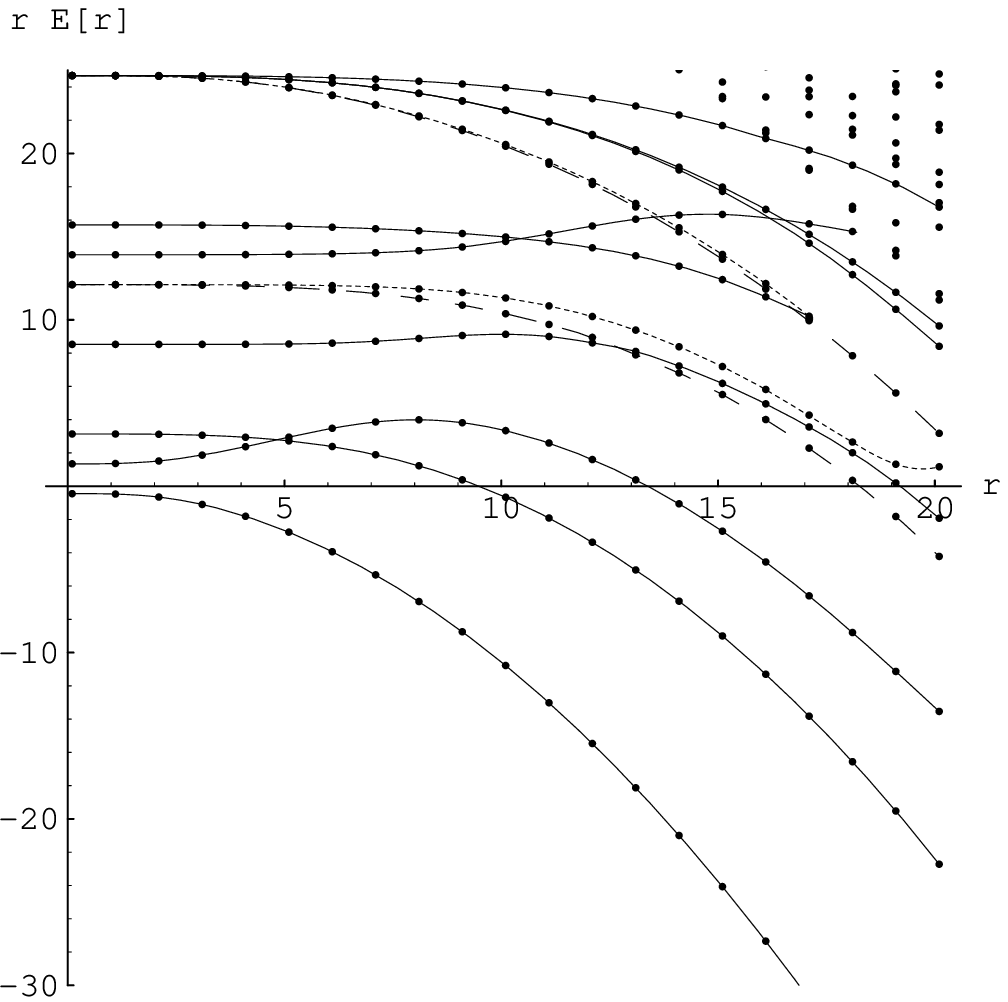} \\
\hbox{(a) Plot of $E(r)$ versus $r$.} &
\hbox{(b) Plot of $r E(r)$ versus $r$.} \\
\end{array}
\]
\caption{The first 13 eigenvalues of $h(r)$ in $\Mm3,14.+\Ph1,5.$.}
\label{fig:314.1}
\end{figure}
}
Some of the lines appear to terminate -- in fact
two real eigenvalues collide and become a complex conjugate pair of
eigenvalues; this is due to truncation effects as explained in
\cite{tcsa1}. 

{}From figure \ref{fig:314.1}a it is possible to estimate the slope of
the ground-state energy and the first three excitation energies in the
scaling region (which we take to be $13 < r < 17$), and we give these 
in table \ref{tab:314} for truncation levels $8, 9\frac 17$ and $10$
to show systematic errors together with estimates of the error of
the measurements, together with the values predicted by the TBA and
perturbation theory.

{
\begin{table}[ht]
\[
\begin{array}{c|c|cccc}
\hbox{Level} & \dim\cH & A  & m_1 &  m_2 & m_3 \\
\hline
 8            & 136 &
        -0.105\pm0.05 & 0.994\pm0.006 & 1.420\pm0.006 & 1.94\pm0.03 \\ 
 9\frac{1}{7} & 280 &
        -0.106\pm0.02 & 1.01\pm0.02~ & 1.417\pm0.002 & 1.95\pm0.03 \\
10            & 285 &
        -0.105\pm0.05 & 0.996\pm0.004 & 1.415\pm0.004 & 1.95\pm0.03 \\
\hline
\hbox{Exact} && -0.1056\ldots & 1 & 1.414\ldots & 1.932\ldots      \\
\end{array}
\]     
\caption{Results for $\Mm3,14.+\Ph1,5.$}
\label{tab:314}
\end{table}
}

However, as we discuss further in section \ref{sec:ad}, 
the particles $A$ and $\bar A$, and $B$, $\bar B$ we used to
diagonalise the \sm\ in section \cite{gt} are not invariant under
${\rm Z}_2$ and consequently there is no direct connection between
these particles and the excitations in the even sector; rather the
particle of lowest mass in the even sector is a symmetric combination
of $A$ and $\bar A$; similarly, the zero-momentum two-particle states
of $A$ and $\bar A$ in the even sector are  $| K(\theta) K(-\theta)
\rangle$  and $|\tilde K(\theta) \tilde K(-\theta) \rangle$. However,
it is still true that the eigenvalues of the \sm\ on these two
particle states are the matrix elements $S_{AA}(\theta)$ and $S_{A\bar
A}(\theta)$, and  we have indicated the first two lines corresponding
to the eigenvalues  by dotted and dashed lines respectively. 

{}From figure \ref{fig:314.1}b we can extract the
\sm\ element $S_{AA}(\theta)$, which we
compare with the conjectured results in figure \ref{fig:314.2}a;
the data is from the {\em second} energy level which corresponds
to these particles, i.e. the 10th eigenvalues of $h(r)$ for
small $r$%
\footnote{
In figs.\ \ref{fig:314.2}a and \ref{fig:314.2}b, small $r$ corresponds
to large $\theta$, and large $r$ to small $\theta$.}. 
As is usual, there is a small discrepancy for small $r$
between the Bethe-Ansatz predictions for $E(r)$ and the actual values
due to finite-size effects, but otherwise the agreement is very good.
In figure \ref{fig:314.2}b we  give the same data for $S_{CC}(\theta)$
which corresponds to the 13th eigenvalue of $h(r)$ for small $r$. 
The results for $S_{BB}$ and $S_{B\bar B}$ are equally convincing.
{
\begin{figure}[htb]
\[
\begin{array}{ll}
\epsfxsize=7.4cm\epsfbox[0 0 288 288]{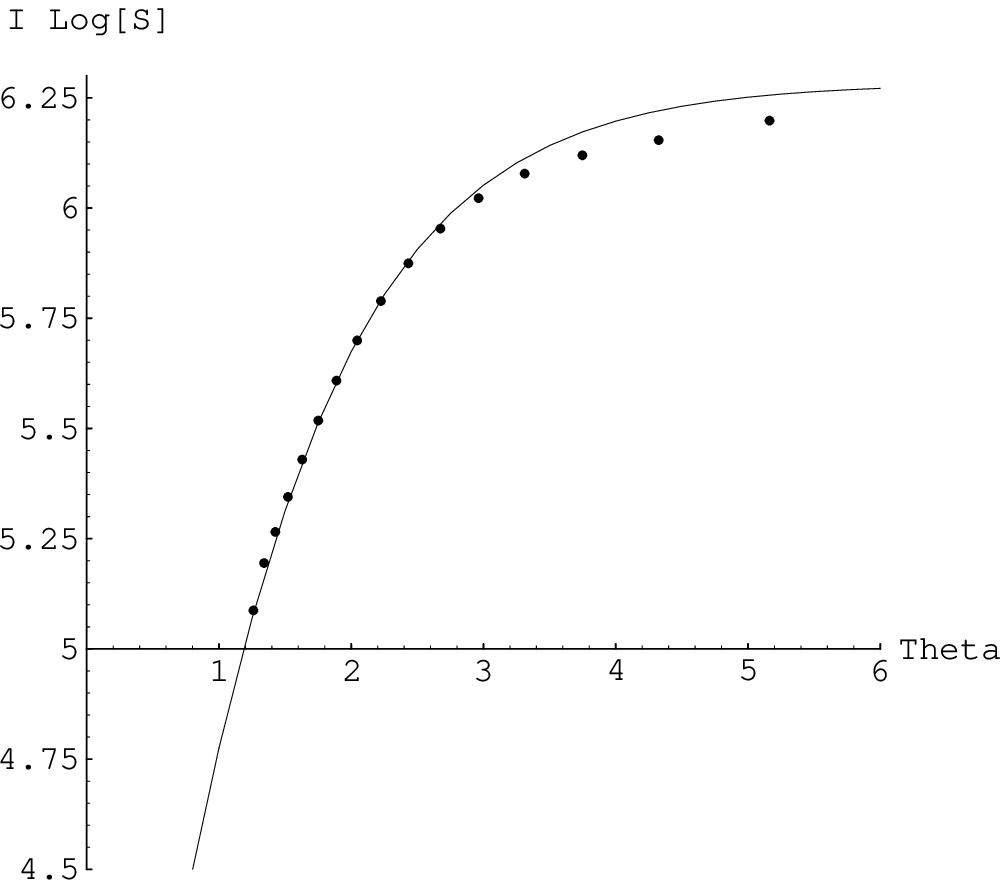} &
\epsfxsize=7.4cm\epsfbox[0 0 288 288]{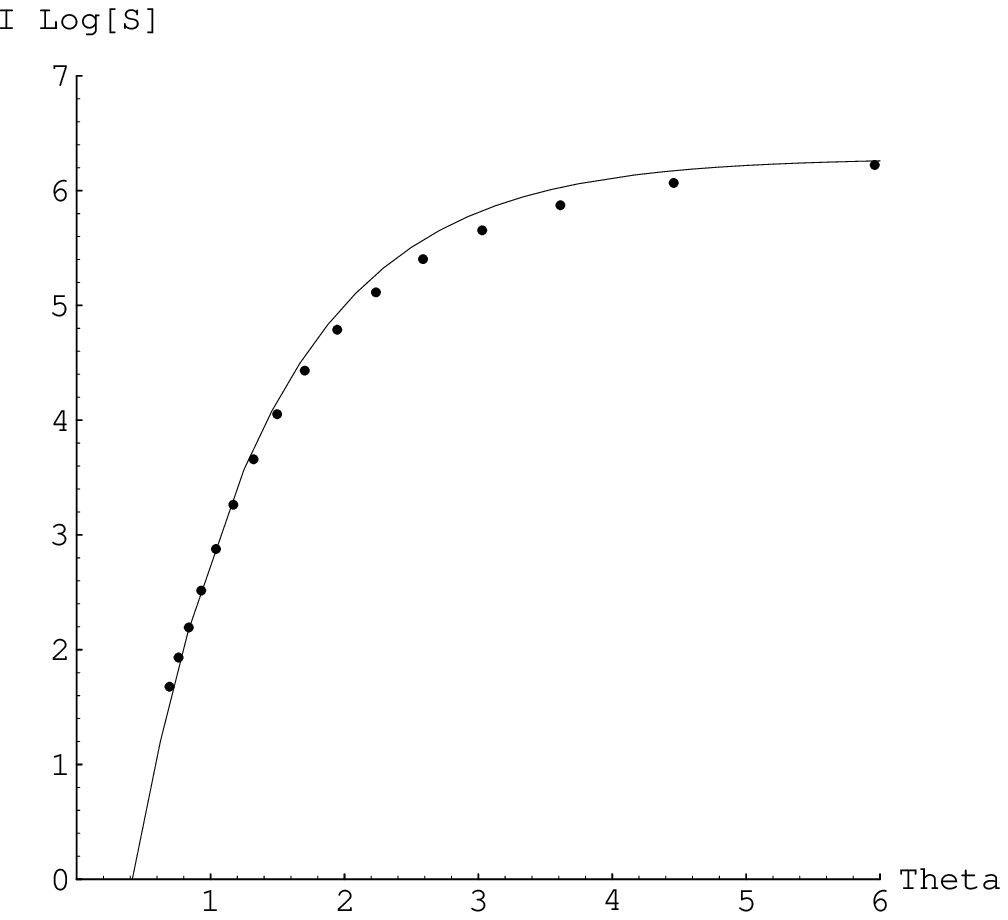} \\
\hbox{(a) The $AA$ \sms.} &
\hbox{(b) The $CC$ \sm.} \\
\end{array}
\]
\caption{The \sms\ as extracted from the TCSA compared with theory in 
$\Mm3,14.+\Ph1,5.$.} 
\label{fig:314.2}
\end{figure}
}

The interpretation of the lowest energy solution of equation
(\ref{eq:ba2})  for $S = S_{A \bar A}$  (the 5th eigenvalue for small
$r$) is extremely interesting. Since $S_{AA}(0)=-1$, the particles $A$
and $\bar A$ are bosons, and so the Bethe Ansatz for the $n$th $A \bar
A$ excitation is 
\begin{equation}
  r\,\sinh(\theta) + \delta_{A\bar A}(2\theta) = 2 n \pi
\ .
\label{eq:aaba}
\end{equation}
Since $\delta_{A \bar A}(\theta)\sim -\theta(3 + 2 \sqrt 2)$ for small
$\theta$ we see that equation (\ref{eq:aaba}) with $n=0$ only has 
solutions for real $\theta$ for $r\leq 6 + 4\sqrt3 = 12.92\ldots$. 
However, for $r$ large, we have identified this 5th line as the single
particle ${\rm Z}_2$ even component of the higher kink doublet 
$B,\bar B$, and hence obtained the estimate of the mass
$m_3=1.95\ldots$. 

We can reconcile these two interpretations when we notice that
for $r> 6+4\sqrt 3$ the solution to equation (\ref{eq:aaba})
corresponds to $\theta$ imaginary. As $r \to \infty$, we see that
$\theta \to i \pi/6$, and there is a pole in $S_{A\bar A}(\theta)$ at
$\theta = i\pi/6$ which corresponds to the particles $B$ and $\bar B$. 
Thus we see direct evidence that the first excited kink (corresponding
to the particles $B$ and $\bar B$) appears as the direct channel pole 
of $K K$ at $\theta=i\pi/6$.

As a consequence, the TCSA gives the value of the \sm\
$S_{A\bar A}(\theta)$ for all real $\theta$ and also for
the imaginary values between $0$ and $i\pi/6$.
In figure \ref{fig:314.3} we plot the theoretical and numerical values
of both $-\delta_{A \bar A}(\theta)$ versus $\theta$ and
$\log(S(\theta))$ versus $-i\theta$. There is excellent
agreement for all real values of $\theta$, and also for 
imaginary values of $\theta$ up to $r\sim17$ (corresponding to
$\theta\sim 0.4i$) where scaling starts to
break down and truncation effects take over%
\footnote{In figs.\ \ref{fig:314.3}a and \ref{fig:314.3}b, small $r$
corresponds to large $\theta$, and $r\to\infty$ to $\theta\to i\pi/6$}.
{
\begin{figure}[htb]
\[
\begin{array}{ll}
\epsfxsize=7.4cm\epsfbox[0 0 288 288]{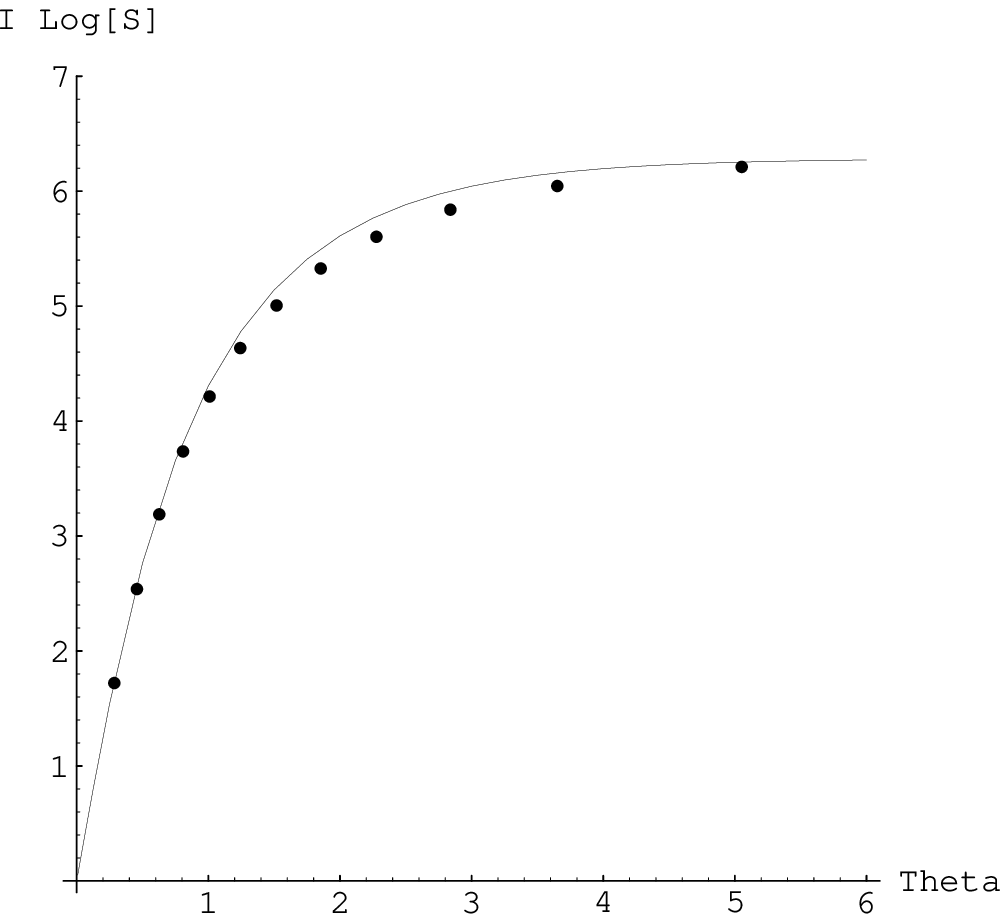} &
\epsfxsize=7.4cm\epsfbox[0 0 288 288]{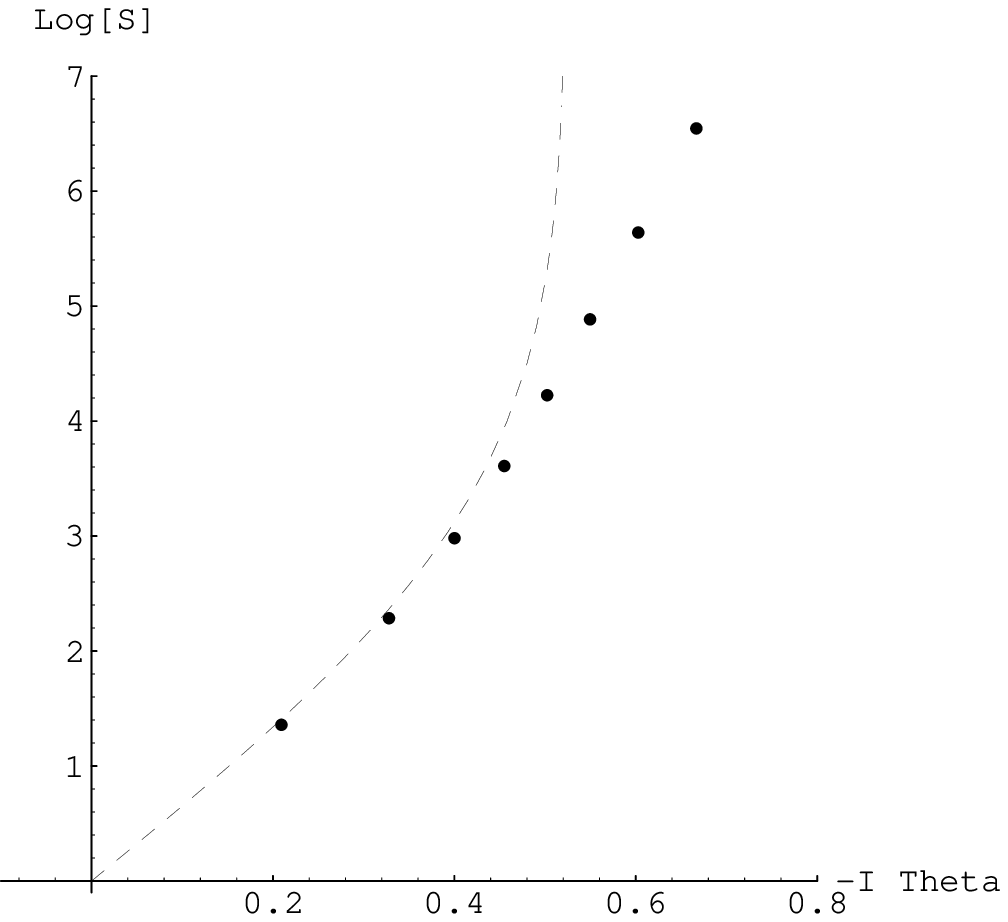} \\
\hbox{(a) $S_{A\bar A}$ for real $\theta$.} &
\hbox{(b) $S_{A\bar A}$ for imaginary $\theta$.} 
\end{array}
\]
\caption{The \sms\ as extracted from the TCSA compared with theory in 
$\Mm3,14.+\Ph1,5.$.} 
\label{fig:314.3}
\end{figure}
}

However, figure \ref{fig:314.1} does not tell the whole story, as we
have only plotted the real eigenvalues. There are also eigenvalues
which are genuinely complex, and in figure \ref{fig:314.4} we show the
real and imaginary parts of those eigenvalues  $\lambda$ amongst the
first 60, for  which  $\Im m(\lambda) > 10^{-6}$. While some of these
are simply the results of truncation error, there are also several
series of lines which are not, and for these we have included the
predictions of the `two-particle BA' using the \sms\ $S_{AB}$ and
$S_{A\bar B}$ and as can be seen, there is excellent agreement.

{
\begin{figure}[htb]
\[
\begin{array}{ll}
\epsfxsize=7.4cm\epsfbox[0 0 288 288]{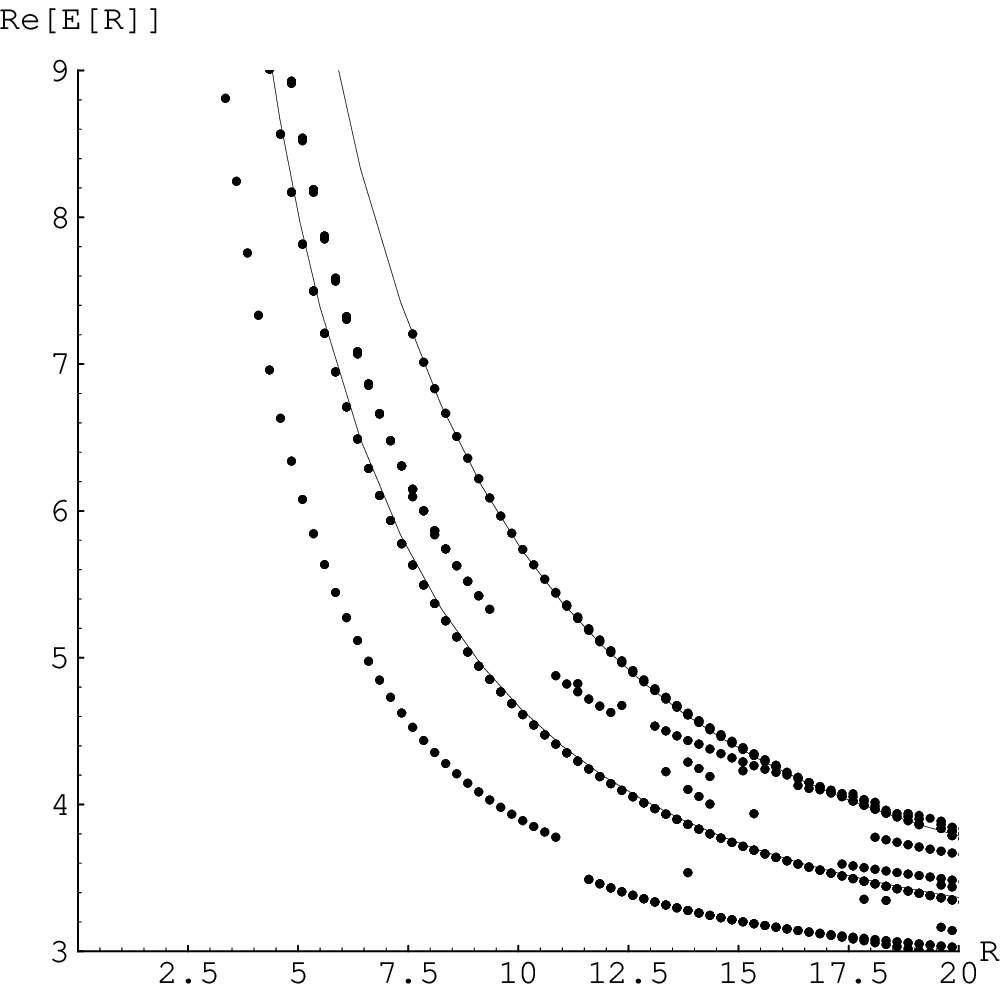} &
\epsfxsize=7.4cm\epsfbox[0 0 288 288]{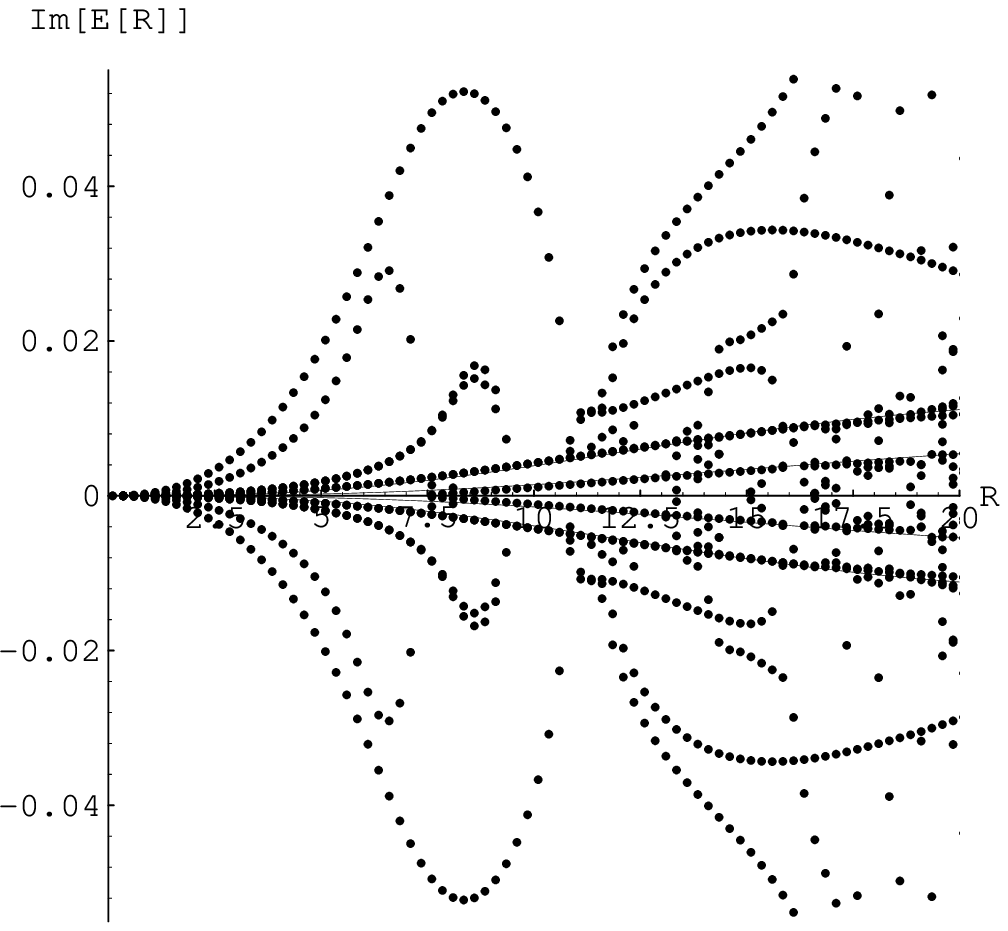} \\
\hbox{(a) $\Re e(\lambda)$.} &
\hbox{(b) $\Im m(\lambda)$.} 
\end{array}
\]
\caption{The complex eigenvalues $h(r)$ in $\Mm3,14.+\Ph1,5.$: 
data and BA predictions}
\label{fig:314.4}
\end{figure}
}

\subsection{The TCSA results for $\Mm3,16.+\Ph1,5.$}

The allowed values of $h$ are $h_{1,i},\ 0<i<16$. It is again possible
to consider several different sectors in which $\Ph1,5.$ acts locally
but which are not themselves necessarily mutually local. In
particular, we can again define three sectors `even', `odd' and
`twisted' for which the fields all have integer spin, with the
following sets of  $\{h_{1,i},\bar h_{1,j}\} \equiv [i,j]$,  
\begin{equation}
\begin{array}{ll}
\hbox{Sector}\ \ &\hbox{Field content} \\
\hbox{even}       & [1,1],\, [3,3],\, [5,5],\, [7,7],\, 
                   [9,9],\, [11,11],\, [13,13],\, [15,15] \\
\hbox{odd}      & [2,2],\, [4,4],\, [6,6],\, [8,8],\, 
                   [10,10],\, [12,12],\, [14,14] \\
\hbox{twisted}   & [2,14],\, [4,12],\, [6,10],\, [8,8],\, 
                   [8,8],\, [10,6],\, [12,4],\, [14,2]. \\
\end{array}
\end{equation}
The coupling constants for first two sectors are those of the `A'
type modular invariant, which we give in appendix \ref{app:cc},
the twisted sector corresponds to the `D' type modular invariant.

To establish that this model has the spectrum we have predicted, it
is again only necessary to consider the `even' sector.
We truncated the space by requiring $L_0 + \bar L_0 \leq 12$ which
leaves 870 states. In figure \ref{fig:316.1} we give the real
eigenvalues of $h(r)$ with the first 14 lines picked out. It is less
clear that scaling has set in for the first five eigenvalues, and
both the masses of the single particles and the \sms\ as extracted
{}from the excited two-particle lines are in good agreement with theory.
However, to achieve a reliable scaling regime it has been necessary to
use a much larger truncated space for this model than the previous
one; indeed when the space is truncated to 285 states the lines
corresponding to the third particle are no longer present, and for
that reason we leave the measurement of its mass blank in table
\ref{tab:316}.
The severity of the truncation effects which still remain can also be
seen in the large gaps in the 2nd, 3rd, 4th and 5th lines (when the
eigenvalues become complex) and which would presumably disappear when
the truncation level is increased.

{
\begin{figure}[ht]
\[
\begin{array}{ll}
\epsfxsize=7.4cm
\epsfbox[0 0 288 288]{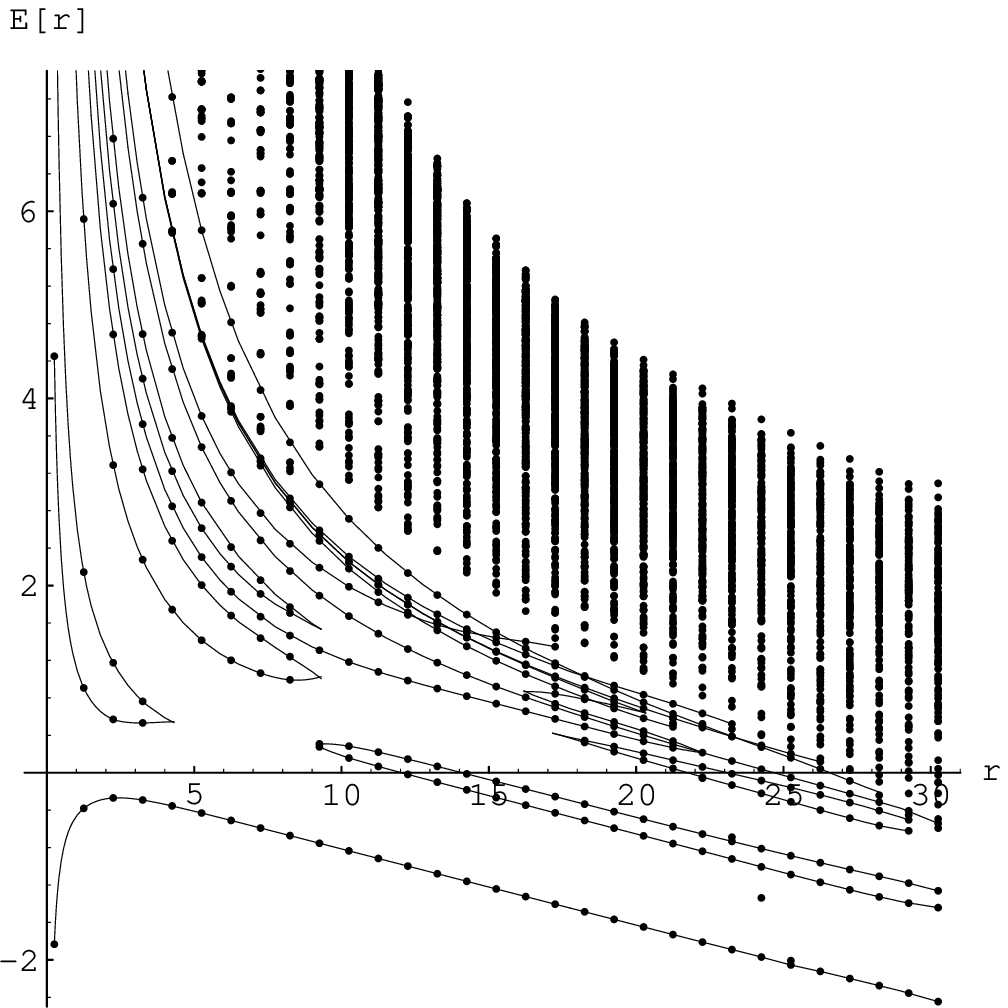} &
\epsfxsize=7.4cm
\epsfbox[0 0 288 288]{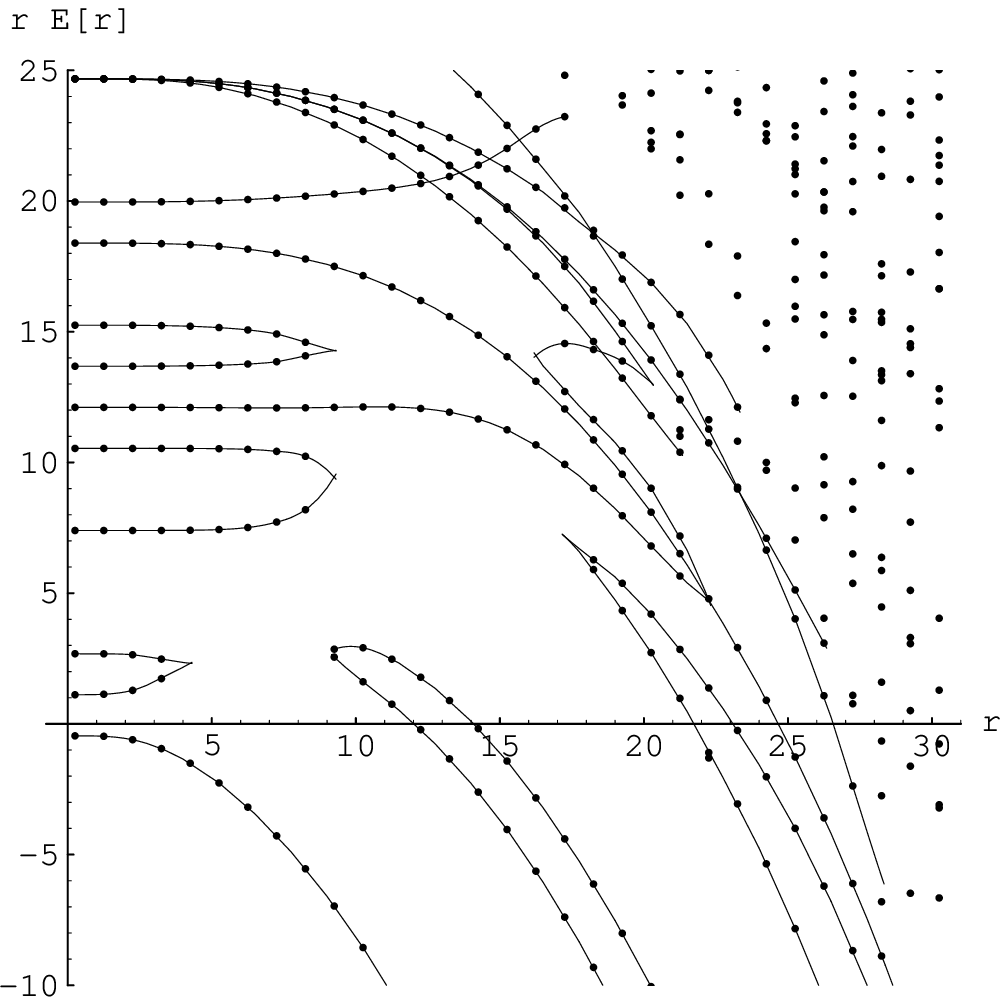} \\
\hbox{(a) Plot of $E(r)$ versus $r$.} &
\hbox{(b) Plot of $r E(r)$ versus $r$.} \\
\end{array}
\]
\caption{The first 14 eigenvalues of $h(r)$ in $\Mm3,16.+\Ph1,5.$.}
\label{fig:316.1}
\end{figure}
}

{}From figure \ref{fig:316.1}a it is possible to estimate the slope of
the ground-state energy and the first three excitation energies in the
scaling region (which we take to be $10 < r < 12$), and we give these
in table \ref{tab:316} for the truncation levels $8\frac34$,
$9\frac34$ and $12$ to
show systematic errors together with estimates of the error of the 
measurements.  

{
\begin{table}[ht]
\[
\begin{array}{c|c|cccccc}
\hbox{Level} & \dim\cH & A  & m_1  & m_2/m_1 & m_3/m_1 \\
\hline
    8\frac34  & 285 & 
         -0.08\pm0.02~ & 1.01\pm0.06 & 
         1.14\pm0.04 & \hbox{---}\\
    9\frac34  & 369 &
        -0.081\pm0.05 & 0.96\pm0.03 & 
         1.175\pm0.02 & 1.81\pm0.02 \\
   12  & 870 &
        -0.0815\pm0.0008 & 0.975\pm0.006 & 
         1.178\pm0.02 & 1.83\pm0.02 \\
\hline
\hbox{Exact} &        & & & 1.175\ldots & 1.827 \ldots \\
\end{array}
\]     
\caption{Results for $\Mm3,16.+\Ph1,5.$}
\label{tab:316}
\end{table}
}

It is clear from the measurement of $A$ that this is not given by the
standard minimal $e_8$ TBA, as should be the case, and that the mass
ratios are consistent with those of the \sms\ of \cite{gt}.

{}From figure \ref{fig:316.1}b we can extract the
eigenvalues of the \sms\ of the fundamental kinks, which we compare
with the conjectured results in figure \ref{fig:316.2}. 
The data points come from the 9th,11th and 23rd eigenvalues of
$h(r)$ (for $r$ small) with the continuous lines being $S_1(\theta)$,
$S_2(\theta)$, and the dashed lines $S_3(\theta)$ and  $S_4(\theta)$.

We can also identify the first two-breather line, which is the 14th
eigenvalue of $h(r)$ for small $R$, and we compare the \sm\ extracted
{}from these numerical results with the theoretical result of
\begin{equation} 
  m_{B_1}
= 2m\sin\frac{\pi}{15}=1.175\ldots m
\ ,
  S_{B_1B_1}
= \ef2,30. \ef10,30. \ef12,30. \ef18,30. \ef20,30. \ef28,30.
\ .
\label{eq:sbb}
\end{equation} 
in figure \ref{fig:316.2}. In this case we have 
$S_{B_1 B_1}(0) = 1$ so that we take $B_1$ fermionic for the
Bethe Ansatz, in accord with the folklore that bosonic and fermionic
particles have $S(0)=-1$ and $1$ respectively; in any case there is
agreement between theory and experiment for this choice.

{
\begin{figure}[htb]
\[
\begin{array}{l}
\epsfxsize=7.4cm
\epsfbox[0 0 288 288]{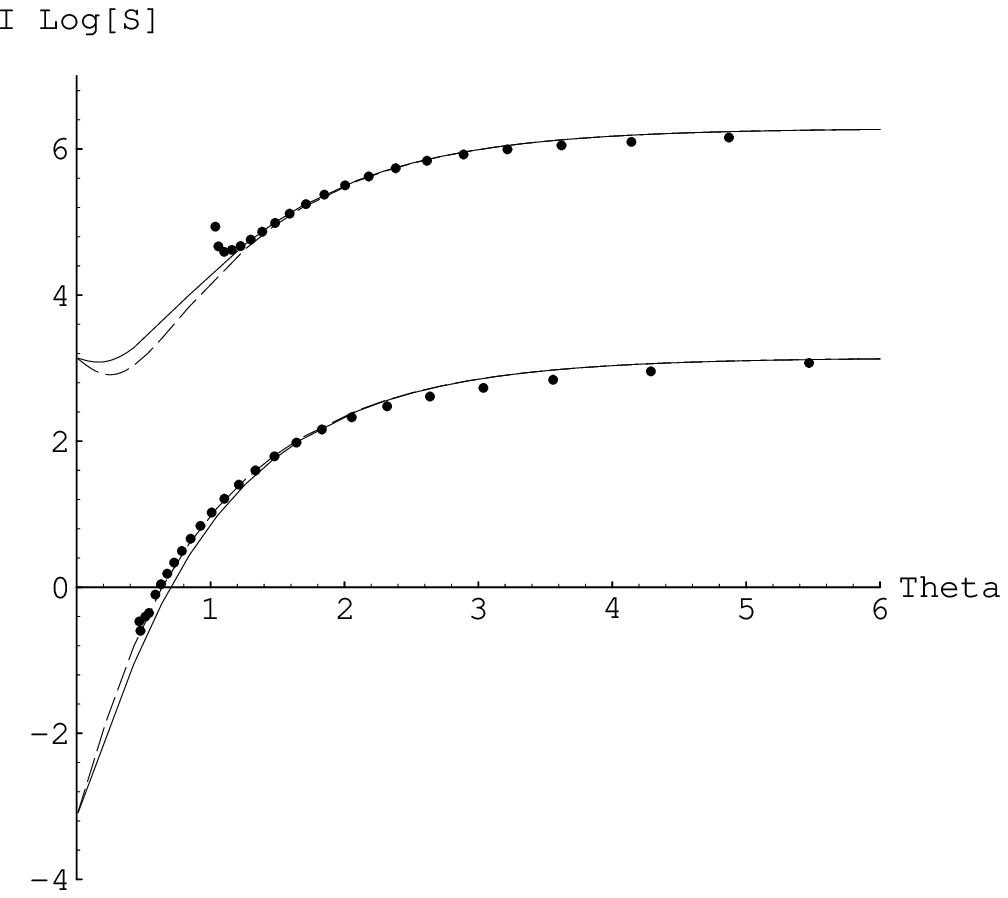} \\
\hbox{(a) The Kink-Kink \sm\ eigenvalues.}
\end{array}
\begin{array}{l}
\epsfxsize=7.4cm
\epsfbox[0 0 288 288]{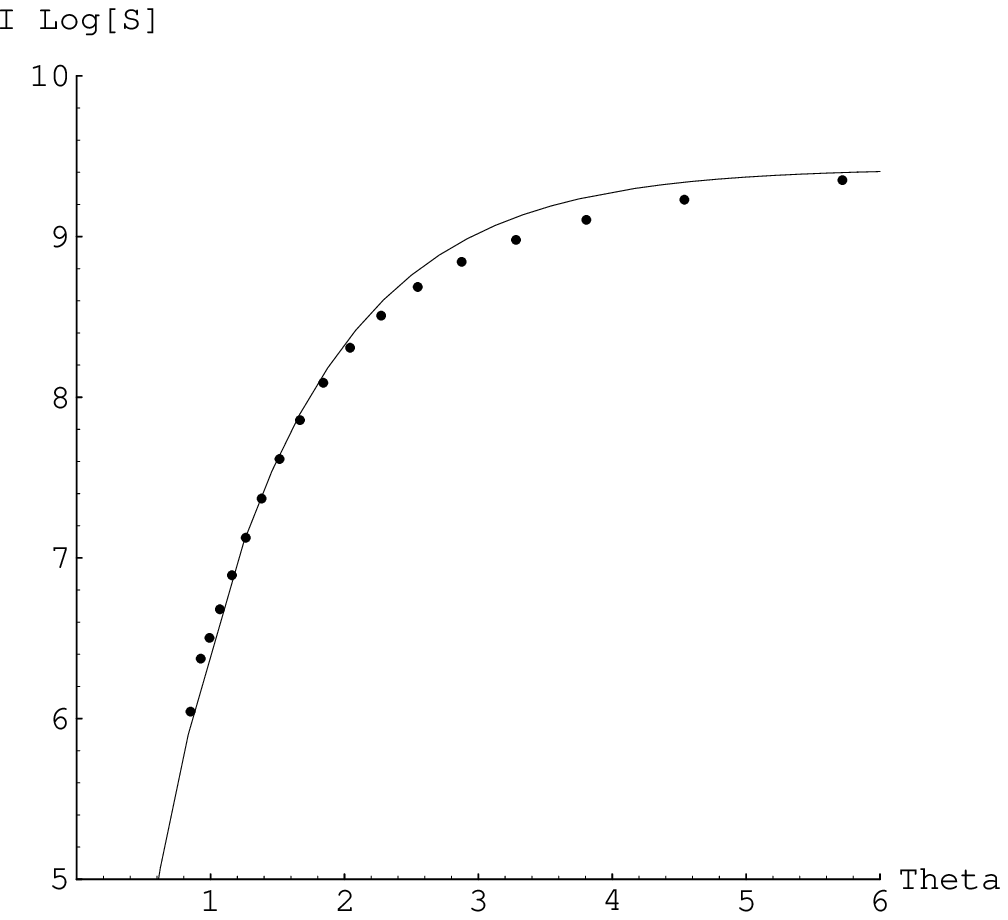} \\
\hbox{(b) The $B_1$--$B_1$ \sm.}
\end{array}
\]
\caption{%
The \sms\ as extracted from the TCSA compared with theory in
$\Mm3,16.+\Ph1,5.$}  
\label{fig:316.2}
\end{figure}
}

%\newpage
\section{${\rm Z}_2$ symmetry and sectors}
\label{sec:ad}

Minimal models of CFT are known to exhibit ${\rm Z}_2$ symmetry, which
has been used above to classify the sectors of the
Hilbert-space. These sectors are preserved under $\Phi_{(1,5)}$
perturbation and therefore it is natural to expect a corresponding
symmetry of the massive field theory. 

There is a natural ${\rm Z}_2$ action on the particles in the
unrestricted ZMS model, completely analogous to the classification of
the irreducible representations of the classical group $SU(2)$ by its
centre ${\rm Z}_2$, with integer spin representations corresponding to
the trivial (even) class and half-integer spin representations to the
nontrivial (odd) class.  
Although the $\cU_q(sl(2))$ action does not survive RSOS restriction,
the ${\rm Z}_2$ action does, and as a consequence, the kinks $K$ are
even and $\tilde K$ odd. 
Since the ${\rm Z}_2$ symmetry is a remnant of the full
quantum group symmetry, it commutes with the \sm, and
it is possible to simultaneously diagonalise both the \sm\ and 
${\rm Z}_2$ on the two-particle states. However, it is {\em not}
possible to do this on one-particle states.  As a result, the
one-particle states which diagonalise the \sm\ (denoted $A,\ {\bar A}$
in case of $\Mm3,14.$ and $\Mm3,16.$ and $A,\ B$ and the case of
$\Mm3,10.$) are {\em not} ${\rm Z}_2$ eigenstates; rather, they are
exchanged with each other under ${\rm Z}_2$. Therefore the natural
description of the content of each sector is in terms of the states
$K$ and $\tilde K$, which are self-conjugate and ${\rm Z}_2$-even or
odd, respectively. 

We now examine the content of the three different sectors (even, odd
and twisted) of the simplest model $\Mm3,10.+\Ph1,5.$, 
as an example. Exactly as for the models $\Mm3,14.$ and $\Mm3,16.$ the
values of  $\{h_{1,i},\bar h_{1,j}\} \equiv [i,j]$ of $\Mm3,10.$ fall
into even, odd and twisted as follows:
\begin{equation}
\begin{array}{ll}
\hbox{Sector}\ \ &\hbox{Field content} \\
\hbox{even}       & [1,1],\, [3,3],\, [5,5],\, [7,7],\, 
                   [9,9],\, \\
\hbox{odd}      & [2,2],\, [4,4],\, [6,6],\, [8,8],\, \\
\hbox{twisted}   & [1,9],\, [3,7],\, [5,5],\, [7,3],\, [9,1].
\end{array}
\end{equation}
The direct sum of the odd and twisted sectors correspond to the
product $(\Mm2,5.+\Ph1,2.)^{\otimes 2}$, and give  
the $D$ type modular invariant in the conformal limit, 
The direct sum of the even and odd sectors gives the $A$-type modular
invariant partition function. Direct TCSA investigation of the odd
sector gives very similar pictures to those of the even sector in the
scaling regime: both sectors have a ground state, a one-particle line
and a very similar structure of two-particle lines. The degeneracy of
the ground state energies is broken by an amount which is
exponentially small for large $r$. 
The picture is very similar for $\Mm3,14.+\Ph1,5.$. 

For the model ${\cal M}_{(3,16)}$, the four eigenvalues of the
kink--kink \sm\ are different, so that we now expect that
the two-particle lines in the even and odd sectors are given by
different Bethe Ansatz results.
In the even sector, we expect the two-particle lines to be linear
combinations of the states 
$| K(\theta) K(-\theta) \rangle $ and
$| \tilde K(\theta) \tilde K(-\theta) \rangle $, whereas those in the even
sector to be linear combinations of 
$| K(\theta) \tilde K(-\theta) \rangle $ and
$| \tilde K(\theta) K(-\theta) \rangle $.
The identification above is consistent with the fact that the
eigenvalues on the ${\rm Z}_2$ even sector, ($S_3$ and $S_4$, the
dashed lines) give a better approximation to the phase-shift extracted
{}from the even sector, than those on the ${\rm Z}_2$ odd sector
($S_1$ and $S_2$, the solid lines).    

This picture can be formulated generically as follows: states
corresponding to RSOS sequences $\{ j_n,\dots j_2,j_1\}$ which start
and end with an even vacuum state (i.e. $j_1$ and $j_n$ are
integer-spin representations, which in our three cases means the state
$| 0\rangle$) originate from the even sector of the underlying CFT,
and the ones starting with an odd vacuum and ending with an odd vacuum
derive from the odd sector.

\section{Character Identities}
\label{sec:chid}

As has been shown extensively, there are interesting sum forms
for Virasoro characters which may be related to the particle
structure of perturbed minimal models (see e.g. \cite{koubek2} for the
cases related to the minimal $e_8$ \sm.). Since the breather sectors of
our related models are the same, and the characters of the
Virasoro representations in these models may be given in terms of sum
formulae, we might expect to find relations between the Virasoro
characters in the two models. This is indeed the case, as we show now.
The character of the irreducible representation $L_{c,h}$ of the
Virasoro algebra of central charge $c$ with highest weight $h$ is
defined as 
\[
  \chi_{c,h}(q) = {\rm Tr}_{L_{c,h}} \left( q^{L_0 - c/24} \right) 
\ .
\]
For pairs $\Mm r,s.$ and $\Mm r',s'.$ of type II ($r'=r/2, s'=2s$) we
find that
\begin{eqnarray}
  \chi_{c_{r,s}, h_{r/2-2m, n}} - 
  \chi_{c_{r,s}, h_{r/2-2m, s-n}} &=& 
  \chi_{c_{r',s'}, h_{m, s'/2-2n}} - 
  \chi_{c_{r',s'}, h_{r'-m, s'/2-2n}}, 
\nonumber\\
  \chi_{c_{r,s}, h_{2m,n}} &=& \chi_{c_{r',s'}, h_{m, 2n}}.
\label{eq:chid}
\end{eqnarray}
Note that these identities do not involve all the characters of either
model. For the pair $\Mm3,10.$ and $\Mm5,6.$ these relations give
\begin{eqnarray}
  \chi_{c_{3,10},h_{1,1}} - \chi_{c_{3,10},h_{1,9}}  &=&
  \chi_{c_{5,6},h_{2,1}}  - \chi_{c_{5,6},h_{3,1}}\ ,\ 
\nonumber\\
  \chi_{c_{3,10},h_{1,3}} - \chi_{c_{3,10},h_{1,7}}  &=& 
  \chi_{c_{5,6},h_{1,1}}  - \chi_{c_{5,6},h_{4,1}}\ ,
\nonumber\\
  \chi_{c_{3,10},h_{1,2}} &=&  \chi_{c_{5,6},h_{1,2}}\ ,\ \
\nonumber\\
  \chi_{c_{3,10},h_{1,4}} &=&  \chi_{c_{5,6},h_{2,2}}\ ,\ \
\nonumber\\
  \chi_{c_{3,10},h_{1,6}} &=&  \chi_{c_{5,6},h_{3,2}}\ ,\ \
\nonumber\\
  \chi_{c_{3,10},h_{1,8}} &=&  \chi_{c_{5,6},h_{4,2}}\ .\ \
\end{eqnarray}
For the pairs $\Mm r,s.$ and $\Mm r,4s.$ of type III we find that
\begin{equation}
  \chi_{c_{r,s}, h_{m,n}} 
= \cases{
  \chi_{c_{r,4s}, h_{(r-m)/2, 2(s-n)}} - 
  \chi_{c_{r,4s}, h_{(r-m)/2, 2(s+n)}} & $m$ odd \cr
  \chi_{c_{r,4s}, h_{m/2, 2n}} - 
  \chi_{c_{r,4s}, h_{m/2, 4s-2n}} & $m$ even \cr}
\ .
\label{eq:chid2}
\end{equation}
In the case of $\Mm3,16.$ and
$\Mm3,4.$ this yields
\begin{eqnarray}
 \chi_{c_{3,4},h_{1,1}} &=& 
 \chi_{c_{3,16},h_{1,6}} - \chi_{c_{3,16},h_{1,10}}\ ,\ 
\nonumber\\
 \chi_{c_{3,4},h_{1,2}} &=& 
 \chi_{c_{3,16},h_{1,4}} - \chi_{c_{3,16},h_{1,12}}\ ,\ 
\nonumber\\
 \chi_{c_{3,4},h_{1,3}} &=& 
 \chi_{c_{3,16},h_{1,2}} - \chi_{c_{3,16},h_{1,14}}\ .\ 
\end{eqnarray}
Analogous relations for type I pairs are obtained by swapping
$r\leftrightarrow s', s\leftrightarrow r'$ and $m\leftrightarrow n$.
All these relations may be proven easily using the formulae in
\cite{RC}.  Whether the identities in equations (\ref{eq:chid}),
(\ref{eq:chid2}) have any real interpretation in terms of the particle
structures of PCFTs is an interesting question to which we hope to
have an answer shortly, although it is interesting to note that in
each case we have investigated when the difference of two characters
appears, it still has an expansion in $q$ with positive coefficients.

%\newpage
\section{Conclusions}
\label{sec:conc}
\vskip -3mm

The above results make it possible to clarify the connection
between $\Phi_{(1,2)}$ and $\Phi_{(1,5)}$ perturbations of minimal
models. We have shown that they can be related through the ZMS
Lagrangian and that the relation falls into one of the three possible
classes of type~I, II or III, according to the relative magnitude of
the effective central charges. Using the S-matrices obtained via RSOS 
restriction in \cite{gt}, we tested them against the predictions of the
TCSA and TBA method. We have confirmed the picture of $\Phi_{(1,5)}$
perturbations based on the RSOS restriction and have shown that the two
related theories are different reductions of the same underlying
affine Toda field theory. The existence of such different restrictions
relies upon the existence of non-equivalent $sl(2)$ subalgebras of the
affine Kac-Moody algebra $a_2^{(2)}$ underlying the ZMS model. It
appears reasonable to expect that such relations should exist between
perturbations of $W$-minimal models corresponding to other imaginary
coupling affine Toda field theories.

The results show that in the type~II case
(models ${\cal M}_{(3,10)}$ and ${\cal M}_{(3,14)}$) we end up with
the same mass spectra, although different scattering theories, whereas
in the cases of type~I (of which the example of $\Mm2,9.$ was treated
in \cite{martins2,martins1}) and type~III (of which $\Mm3,16.$ is an
example) even the mass spectra are different just as expected. In the
two type~II cases the TBA equations turned out to be the same as those
of the corresponding $\Ph1,2.$-perturbed conformal field theories.

We have also shown that the strange features of the \sms\ for
$\Mm3,14.+\Ph1,5.$ found in \cite{gt} are in fact correct. This model
does indeed break unitarity much more severely than the $\Ph1,2.$ and
$\Ph1,3.$ perturbations considered before. The spectrum is not
entirely real, and it is well described by the \sms\ no longer being
pure phases. We believe that this is a generic feature of RSOS
restrictions, and that the previous situation with a real spectrum is
in fact rather exceptional. Indeed, looking at the \sms\ found for
$a_2^{(1)}$ Toda theory in \cite{GGan1} we see that the \sms\ for
soliton--breather scattering (which should not be altered by RSOS
reduction) are also not pure phases in general.

We have also shown that in one case, the Bethe Ansatz equations give a
good approximation for a single-particle state energy by continuation
to imaginary rapidity. While it is easy to check that in general this
is not the case, it is true that the leading corrections to the
particle mass in $\Mm2,5.+\Ph1,2.$ are also given by the BA equations,
and that this can be seen by examining the large $R$ behaviour of the
full TBA equations describing this energy \cite{fullTBA}. We believe
that in our case, the full TBA equations are simply dominated by the
BA equations for all values of $R$, rather than simply for large $R$.

Unfortunately, due to our lack of knowledge about the
conformal field theory three-point coupling constants for $D$-type
modular invariants of non-unitary models, we have not been able to 
extract precise TCSA predictions for the twisted sector, although
preliminary results (using some guesses for the structure constants)
have turned out to be in good agreement with our expectations based on
the $\rm Z_2$ picture. It seems worthwhile to tackle the problem of
the coupling constants in the non-unitary minimal models and in this
way to facilitate the further investigation of their structure as
well.

Since the RSOS restriction does not modify the breather sector of the
affine Toda field theory, one can expect that this sector is identical
in corresponding $\Phi_{(1,2)}$ and $\Phi_{(1,5)}$ perturbations
irrespective of the type~I, II or III nature of the pair. This leads
to the existence of `strange' character identities
between characters of different minimal models. In addition to that,
in the type~II case we expect relations between the states
corresponding to the solitonic sectors as well. Some character
identities of this type have been described, although it is not clear
whether they originate from some connections between different PCFTs 
in the case of general minimal model pairs.

\vskip 3mm
\centerline{\bf Acknowledgements}

We would like to thank W.~Eholzer, J.M.~Evans,
A.~Honecker, A.~Koubek, N.J.~MacKay, G.~Mussardo and R.A.~Weston for
useful  discussions and comments at various stages.
We would especially like to thank P.~Dorey for explanations of the
results in \cite{fullTBA} and E.~Corrigan for his suggestion that we
try again to complete the bootstrap for our models.
GMTW is supported by an EPSRC advanced fellowship.

Much of this work was done while the authors were in DAMTP, Cambridge,
where HGK was supported by a research fellowship from Sidney Sussex
College, Cambridge, and GT by the Cambridge Overseas Trust.
We also acknowledge partial support from PPARC.
The numerical work was performed in MATHEMATICA and MAPLE on equipment
supplied under SERC grant GR/J73322.

%\newpage
\appendix
\section{Coupling constants in non-unitary conformal field theories} 
\label{app:cc}

The coupling constants for fields of type $\Ph1,a.$ are given in
\cite{DFat} in an asymmetric fashion: they have two sets of fields
$\varphi_a, \bar\varphi_a$  which for our purposes are just
(different) multiples of $\Ph1,a.$
Using the results of \cite{DFat} we arrive at
\[
  \left\langle \varphi_a \right|\,
  \varphi_b(1)\,
  \left| \varphi_c \right\rangle
= \tilde C_{abc}
\ ,\ \ 
  \left\langle \varphi_a \right|\left.
               \varphi_b \right\rangle
= \tilde C_{ab1}
\]
where the coupling constants are 
\begin{equation}
  \tilde C_{abc}
= \frac{ \xi( { \frac{b+c-a-1}2 } ) }{\xi( a-1 )}
  \frac{ \xi( { \frac{a+c-b-1}2 } ) }{\xi( b-1 )}
  \frac{ \xi( { \frac{a+b-c-1}2 } ) }{\xi( c-1 )}
  \frac{ \eta({ \frac{a+b+c-1}2 } ) }{\eta( 1  )}
\ ,
\label{eq:ctilde}
\end{equation}
where
\[ 
  \xi(a)  = \prod_{j=1}^a \gamma( \rho j )
\ , \ \
  \eta(a) = \prod_{j=1}^a \gamma( \rho j - 1)
\ , \ \
  \gamma(x) = \Gamma(x)/\Gamma(1-x)
\ , \ \
  \rho    = r/s
\ .
\]
{}From the unnormalised coupling constants $\tilde C_{abc}$ it is then
easy to choose normalised fields 
$\Ph1,a. = \frac{1}{\alpha_a} \varphi_a$ such that 
$\left\langle \Ph1,a. \right|\, \Ph1,b.(1)\,
 \left| \Ph1,c. \right\rangle$ is real and 
$  \left\langle \Ph1,a. \right|\left. \Ph1,a. \right\rangle
= \pm 1 $.

We do not give the coupling constants for the D type modular invariant
involving the fields in the twisted sector as these are not to be
found in \cite{DFat}. Petkova and Zuber \cite{PZub} give these for the
unitary minimal models, and it would be interesting to extend their
results to all minimal models, but we do not do that here.

\end{document}